\documentclass[apj]{emulateapj}

\shorttitle{Dense Molecular Gas and Stars at Circumnuclear Starburst Ring}
\shortauthors{Pan et al.}

\usepackage[]{natbib}

\begin{document}

\title{Formation of Dense Molecular Gas and Stars at the Circumnuclear Starburst Ring in the Barred Galaxy NGC~7552}

\author{Hsi-An Pan\altaffilmark{1,2,3}}
\author{Jeremy Lim\altaffilmark{4,7}}
\author{Satoki Matsushita\altaffilmark{3}}
\author{Tony Wong\altaffilmark{5}}
\author{Stuart Ryder\altaffilmark{6}}
\email{pan.h.a@nao.ac.jp}
\affil{$^{1}$Department of Astronomical Science, The Graduate University for Advanced Studies, Shonan Village, Hayama, Kanagawa 240-0193, Japan}
\affil{$^{2}$Nobeyama Radio Observatory, NAOJ, Minamimaki, Minamisaku, Nagano 384-1305, Japan}
\affil{$^{3}$Academia Sinica, Institute of Astronomy \& Astrophysics (ASIAA), P.O. Box 23-141, Taipei 10617, Taiwan}
\affil{$^{4}$Department of Physics, University of Hong Kong, Pokfulam, Hong Kong}
\affil{$^{5}$Department of Astronomy, University of Illinois, 1002 W. Green Street, Urbana, IL 61801, USA}
\affil{$^{6}$Australian Astronomical Observatory, P.O. Box 915, North Ryde, NSW 1670, Australia}

\altaffiltext{7}{Adjunct Research Fellow, Academia Sinica, Institute of Astronomy \& Astrophysics (ASIAA), P.O. Box 23-141, Taipei 10617, Taiwan}

\begin{abstract}
We present millimeter molecular-line complemented by optical observations, along with a reanalysis of archival centimeter \ion{H}{1} and continuum data, to infer the global dynamics and determine where dense molecular gas and massive stars preferentially form in the circumnuclear starburst ring of the barred-spiral galaxy NGC~7552.   We find diffuse molecular gas in a pair of dust lanes each running along the large-scale galactic bar, as well as in the circumnuclear starburst ring.  We do not detect dense molecular gas in the dust lanes, but find such gas concentrated in two knots where the dust lanes make contact with the circumnuclear starburst ring.  When convolved to the same angular resolution as the images in dense gas, the radio continuum emission of the circumnuclear starburst ring also exhibits two knots, each lying downstream of an adjacent knot in dense gas. The results agree qualitatively with the idea that massive stars form from dense gas at the contact points, where diffuse gas is channeled into the ring along the dust lanes, and later explode as supernovae downstream of the contact points.  Based on the inferred rotation curve, however, the propagation time between the respective pairs of dense gas and centimeter continuum knots is about an order of magnitude shorter than the lifetimes of OB stars.  We discuss possible reasons of this discrepancy, and conclude that either the initial mass function is top-heavy or massive stars in the ring do not form exclusively at the contact points where dense molecular gas is concentrated.

\end{abstract}
\section{Introduction}
\label{sec_intro}
Over half \cite[e.g., $\gtrsim$70\%;][]{Esk00} of all spiral galaxies in the Local Universe exhibit large-scale bars.  Such barred galaxies exhibit circumnuclear starbursts more frequently than non-barred spiral galaxies \citep{Jog05,Ell11}.  The circumnuclear starburst usually takes the form of a ring centered on the nucleus \citep[e.g., review by][]{But96}, and has a star-formation rate (SFR) that can make up a significant fraction, if not dominate, the total SFR of the galaxy \citep{Jog05}.  Is gas transported from the surrounding regions towards the center of the galaxy to fuel the circumnuclear starburst, and if so how?  If bars transport gas inwards, then why we do not see circumnuclear starbursts in virtually all barred spiral galaxies?  Stated in a different way, why is star formation is strongly promoted in the circumnuclear region of some galaxies?  In this work, we address the last question through observations of the barred spiral galaxy NGC~7552, which exhibits a circumnuclear starburst ring with a SFR of 10--15 M$_{\sun}$ yr$^{-1}$ that comprises at least half of the total SFR of this galaxy.

Theoretical studies suggest that, in galactic disks, gravitational torques induced by bar potentials drive gas inflows.  In this scenario, molecular gas experiences two principal shocks: (i) at the two leading edges of the large-scale bar, where dust lanes (indicating, presumably, the pileup of molecular gas) are commonly seen \citep[e.g.,][]{She02}; (ii) at the two locations where the inner terminal portions of the bar make contact with the circumnuclear starburst ring, where twin peaks in molecular gas as traced by CO are commonly observed \citep[e.g.,][]{Ken92,Sak99}.
During the passage of molecular gas in the galactic disk through the bar, angular momentum is transferred from the gas to the bar through the aforementioned shocks, resulting in a net inflow of gas along the dust lanes.  The inflowing gas is eventually halted at an inner Lindblad resonance (ILR; or, if there are two ILRs, then between the inner ILR and outer ILR), which has the form of a ring, where the bar terminates and within which the net torque exerted by the bar is zero \citep{But96}.  Because the formation of a circumnuclear gas ring requires a bar that is strong enough (i.e., to induce a sufficiently strong non-axisymmetric potential) to promote adequate gas infall, the mere presence of a bar is not sufficient for the presence of a circumnuclear gas ring.  Simulations \citep[e.g.,][]{Reg04}, as well as observations \citep[e.g.,][]{Sak99,Lau07}, suggest that galaxies with long and massive (morphological classification SBa) bars are more effective at channeling gas towards their centers than those with weak bars (SBc), and hence preferentially host circumnuclear gas rings.

Although molecular gas has been observed to accumulate along the leading edges of the bar and at the ILR in agreement with model predictions, there is a striking difference between the observed properties of the molecular gas and the star-formation activity in these two regions.  The molecular gas in (at the leading edge of) the bar has been found to be much more diffuse on average \citep[e.g.][]{Dow96, Hut00, Sch02}, and the star-formation activity in the bar also much less pronounced, than that in the circumnuclear starburst ring.  Based on observations of NGC~6951, a strongly barred spiral galaxy that exhibits a circumnuclear starburst ring and which in many other respects are remarkably similar to NGC~7552 as explained below,
 \citet{Koh99} described how diffuse molecular gas may collect in the circumnuclear starburst ring of this galaxy and the circumstances under which this gas subsequently becomes sufficiently dense to form stars.  They found that, just like in a number of other barred galaxies having circumnuclear starburst rings, diffuse molecular gas as traced in $^{12}$CO is concentrated in two knots on opposite sides of the ring where a pair of dust lanes, each running along the (leading edge of the) bar on opposite sides of the nucleus, make contact with the ring, demonstrating that the twin CO peaks likely represent molecular gas channeled inwards along the dust lanes.   
For the first time, they showed that dense molecular gas as traced in HCN is not cospatial with the CO knots, but is instead concentrated in two knots located just downstream of the respective CO knots.  \citet{Koh99} argued that diffuse molecular gas, channeled inwards along $x_1$ orbits (closely traced by the dust lanes), is too turbulent to become dense where they intersect with $x_2$ orbits (traced by the circumnuclear starburst ring), as the gas at these locations (delineated by the CO knots) experience relatively strong shocks.  Instead, they proposed that the molecular gas only becomes dense further downstream where the aforementioned turbulence has partially dissipated, leading to the collapse of gas clouds via gravitational (Toomre-type) instabilities \citep{Too64} and subsequent star formation.

Because intercomparisons can provide insights not possible from studies of a single object alone, here we test the idea proposed by \citet{Koh99} on  another galaxy.  NGC~7552 is a member of the Southern Grus Quartet at a distance of 22.3~Mpc \cite[][so that 1$\arcsec$ = 107 pc]{Ken03}.  Observations of the entire group in atomic hydrogen (\ion{H}{1}) gas have revealed an extension on the south-western side of NGC~7552 indicative of tidal interactions with one or more group members \citep{Dah05}, although no such distortion is visible in the optical disk suggesting that only the outskirt of NGC~7552 is strongly perturbed.  The gross optical morphology of NGC~7552 --- circumnuclear starburst ring, large-scale bar with dust lanes along the leading edges, and two prominent spiral arms emerging from the ends of the bar as can be seen in Figure~\ref{FIG_Optical} --- is remarkably similar to that of NGC~6951, including even the orientation of the abovementioned features in the sky.   The large-scale stellar disk of NGC~7552 is close to face-on, with an axial ratio between the optical major and minor axes indicating an inclination of about $25\degr$ (for comparison, also based on axial ratios from optical images, the LEDA extragalactic database quotes an average inclination of 23$\degr$, and \citet{Fei90} infer an inclination of 28$\degr$).  The nuclear activity of NGC 7552 is classified as a Low-Ionization Nuclear Emission-line Region (LINER) with \ion{H}{2} regions \citep{Dur87}.  \citet{Cla92} find that the molecular gas in NGC~7752 as traced in $^{12}$CO~(1--0) is strongly concentrated in the nucleus and the bar.

\citet{Fo94a} have imaged the circumnuclear starburst ring of NGC~7552 in the radio continuum at 6~cm and 3~cm, providing the least-obscured view of the ring so far available.  As explained in greater detail later, the ring is nearly circular, and exhibits five prominent knots.  The radio spectrum of each knot falls between 6~cm and 3~cm, indicating nonthermal synchrotron emission produced, presumably, as a result of supernova explosions.  \citet{Fo94b} also imaged the ring in H$\alpha$ and \citet{Sch97} in Br$\gamma$, where both these lines trace gas ionized by recently-formed massive stars.

Our work presented here involves multi-line radio interferometric complemented by optical observations of NGC 7552, along with a reanalysis of archival data for this galaxy in \ion{H}{1} and centimeter continuum. In $\S$\ref{sec_obs} we provide details on the observations performed and the data reanalyzed, and in $\S$\ref{sec_result} the results obtained.
In $\S$\ref{sec_physical_properties} we derive the physical properties of the molecular gas, and in $\S$\ref{sec_rotation_curve_kinematics} the global kinematics of NGC~7552.  Then, in $\S$\ref{sec_sf}, we examine the implications of our results and analyses for the formation of dense molecular gas and stars in the circumnuclear starburst ring of NGC~7552, and propose an important revision to the picture advanced by \citet{Koh99}, based on observations of the circumnuclear starburst ring in NGC~6591, that is better able to explain the observed properties of the circumnuclear starburst ring in NGC~7552 as well as more recent observations of dense gas in the circumnuclear starburst ring of NGC~6951.  Finally, in $\S$\ref{sec_summary}, we provide a concise summary of our work.

\section{Observations and Data Reduction}
\label{sec_obs}
Our millimeter molecular-line observations toward the center of NGC~7552 focussed on four bright tracers of molecular hydrogen gas, $^{12}$CO, $^{13}$CO, HCN, and HCO$^{+}$.  $^{12}$CO traces relatively diffuse ($\sim$$10^{2-3} {\rm \ cm^{-3}}$) molecular gas characteristic of the bulk of giant molecular clouds (GMCs) in our galaxy, whereas HCN and HCO$^{+}$ traces relatively dense ($\sim$$10^{5-6} {\rm \ cm^{-3}}$) molecular gas characteristic of dense condensations in GMCs where stars form; as we will show in $\S$\ref{sec_lineratio_LVG}, in our observations $^{13}$CO traces molecular gas at intermediate densities.  The spatial distribution of $^{12}$CO therefore indicates locations where atomic hydrogen gas has presumably been compressed to form molecular hydrogen gas, whereas the spatial distribution of HCN and HCO$^{+}$ indicates locations where the molecular gas has attained densities sufficiently high to form stars.  These molecular-line observations also provide information on the gas kinematics: as described in $\S$\ref{subsec_brandt_rotation_curve}, we used the $^{12}$CO image to infer a rotation curve for the central region of NGC~7552.  The \ion{H}{1} image of NGC~7552 provides information on the gas kinematics beyond the central region imaged in molecular gas, and as described in $\S$\ref{subsec_brandt_rotation_curve} was used to infer a rotation curve for the galaxy between the innermost to the outermost detectable regions of the \ion{H}{1} disk.  The rotation curve thus inferred was used to derive the predicted locations of dynamical resonances in the galaxy, as well as to compute dynamical timescales in the circumnuclear starburst ring.  We compared the predicted locations of dynamical resonances to individual features in NGC~7552 from our wide- and narrow-band optical observations, thus providing an assessment of the reliability of the inferred rotation curve.  The narrow-band H$\alpha$ image from our optical observations also provided estimates of the SFRs in different parts of the galaxy.  Finally, from the radio continuum image of the circumnuclear starburst ring, we inferred locations where large concentrations of massive stars recently met their demise as supernovae.

\subsection{Optical observations: Gemini South}
\label{sec_gemini}
The optical observations of NGC~7552 reported here are the product of a winning entry in the 2011 Australian Gemini School Astronomy Contest, for which one of us (S. Ryder) was a co-investigator on the contest proposal.  Images of the galaxy in $g^{\prime}$, $r^{\prime}$, and $i^{\prime}$ filters were obtained with the Gemini Multi-Object Spectrograph \citep[GMOS;][]{gmos04} attached to the Gemini South Telescope as part of the program GS-2011A-Q-12 (PI: C. Onken). The filters are similar to the filters used by the Sloan Digital Sky Survey (SDSS). We also obtained an H$\alpha$ image with a narrow-band H$\alpha$ filter centered on 656 nm with a 7 nm wide bandpass. The observations were done in seeing between $0\farcs6-0\farcs8$ under photometric conditions on 2011 July 23 UT. 
Four exposures per filter were obtained with integration time between 100 and 300~s, with each exposure offset by 10~arcsec spatially to allow filling-in of the inter-CCD gaps in GMOS. On-chip binning yielded a pixel scale of $0\farcs146$ pix$^{-1}$.

The data were reduced and combined using the {\em gemini} package (V1.10) within {\sc iraf}\footnote{{\sc iraf} is distributed by the National Optical Astronomy Observatories, which are operated by the Association of Universities for Research in Astronomy, Inc., under cooperative agreement with the National Science Foundation.}. A master bias frame (constructed by averaging with 3-$\sigma$ clipping a series of bias frames) was subtracted from all raw images. Images of the twilight sky were used to flatfield the images, then the dithered galaxy images were registered and averaged together with the {\em imcoadd} task to eliminate the inter-CCD gaps, bad pixels, and cosmic rays.

\subsection{Radio observations: Australia Telescope Compact Array}
\label{sec_atcaobs}
\subsubsection{\ion{H}{1}}
\label{subsec_atcaobs_hi}
We re-created the 21-cm neutral atomic hydrogen (\ion{H}{1}) maps of \citet{Dah05}, who kindly provided us with calibrated data, using robust weighting 
to provide a higher angular resolution than the maps he published.  The parameters of the \ion{H}{1} observation are summarized in Table \ref{TAB_obspara}.   The \ion{H}{1} maps presented here have a synthesized beam of $20\farcs0 \times 20\farcs0$ at full-width half-maximum (FWHM) (by comparison, the \ion{H}{1} maps published by \citet{Dah05} have a synthesized beam of $34\farcs1 \times 30\farcs0$).  In our channel maps, the root-mean-square (rms) noise level in each $13 {\rm \ km \ s^{-1}}$ channel is 1.1~mJy beam$^{-1}$.  The integrated \ion{H}{1} intensity and intensity-weighted mean \ion{H}{1} velocity maps, as well as all other such maps presented in this work, were constructed using the task MOMNT in the AIPS package; this task allows us to select or discard pixels in each channel map based on whether their corresponding intensities when smoothed over a selected number of channels exceed a specified threshold.  The \ion{H}{1} disk of NGC~7552 is much smaller than the size of the primary beam, and hence no primary beam correction was applied to the data.

\subsubsection{Radio Continuum}
\label{subsec_atcaobs_continuum}
We re-processed the radio continuum data taken by \citet{Fo94a} at 3~cm and 6~cm, and re-created their maps to permit our own analyses as well as provide digital versions for use in making overlays.  As summarized in Table \ref{TAB_obspara}, the observations of \citet{Fo94a} were conducted  in one 12-hour synthesis on 11--12 February 1993 using the Australia Telescope Compact Array (ATCA) in its 6C configuration.  This configuration provides baseline lengths ranging from 153~m to 6000~m, making it possible to image structures (with uniform brightnesses) as large as about 285$\farcs$0.  The primary flux calibrator used was PKS~B1934--638, for which we adopted a flux density of 5.83~Jy at 6~cm (4786~MHz) and 2.84~Jy at 3~cm (8640~MHz).  The quasar PKS~B2311--452 was used for complex gain calibration.  

We calibrated the data in the standard manner using the MIRIAD package.  To improve the dynamic range of our maps, we self-calibrated the datasets at both 6~cm and 3~cm from the maps obtained at the respective wavelengths.  The final map at 6~cm has a synthesized beam of $2\farcs0\times1\farcs3$ at a position angle (PA) of 0$\degr$, and an rms noise fluctuation of $72 {\rm \ \mu Jy \ beam^{-1}}$.   The final map at 3~cm has a synthesized beam of $1\farcs1 \times 1\farcs1$ and an rms noise level of $66 {\rm \ \mu Jy \ beam^{-1}}$.  Both the angular resolutions and noise levels of our maps are, of course, similar to those in the maps shown by \citet{Fo94a}.  Given the small angular extent of the circumnuclear radio continuum emission from NGC\,7552, no primary beam correction was applied to the data.

\subsubsection{HCN and HCO$^{+}$}
\label{subsec_atcaobs_hcn}
We observed NGC~7552 in HCN~(1--0) and HCO$^{+}$~(1--0) with the ATCA on three separate occasions between May and September 2005 in the H75, H168, and H214 configurations.  As summarized in Table \ref{TAB_obspara}, this combination of array configurations provided baseline lengths ranging from 30~m to 200~m, and hence sensitivity to structures (with uniform brightnesses) as large as about 23\arcsec.  Five of the six antennas of the ATCA were equipped with dual polarization receivers in the 3~mm band, covering a frequency range from 85~GHz to 105~GHz.   At the wavelength of the HCN~(1--0) (rest frequency of 88.6318470 GHz) and HCO$^{+}$~(1--0) (rest frequency of 89.1885230 GHz) lines, the primary beam of the antennas is $\sim$36\arcsec.  We configured the correlator to observe the HCN~(1--0) and HCO$^{+}$~(1--0) lines simultaneously in two spectral windows, with the individual windows covered by 64 channels of width 2~MHz ($7 {\rm \ km \ s^{-1}}$) each.  Undersampling in the lag domain makes the spectral channels not completely independent; we therefore binned the data into $15 {\rm \ km \ s^{-1}}$ channels for making maps.  We adopted a pointing center of $\rm \alpha_{2000} = 23^{\rm h}16^{\rm m}11^{\rm s}$ and $\rm \delta_{2000} = -42\arcdeg35\arcmin05\arcsec$, coinciding closely with the center of the circumnuclear starburst ring.  The typical single-sideband system temperatures varied over the range 250--300 K, and the total on-source integration time was about 17~hr.

We reduced the data in the standard manner using MIRIAD, a procedure that involved baseline corrections, flagging of bad channels and shadowed baselines, calibration, and finally imaging.  The absolute flux density scale was set by Uranus or Mars, excluding baselines for which the planets were heavily resolved.  We used the bright quasar PKS~B1921--293 for bandpass calibration, and the quasar PKS~B2326--477 (lying within 5\arcdeg\ of our target) for complex gain calibration.  We also observed another quasar, PKS~B2255--282 (lying within 15\arcdeg\ of our target), at frequent intervals to verify the phase solutions.  

We made channel maps using robust weighting, yielding a synthesized beam of 2\farcs6 $\times$ 2\farcs0, at a PA of 80\arcdeg.  
The final maps have an rms noise level of 4.5~mJy in HCN~(1--0) and 4.2~mJy in HCO$^{+}$~(1--0) for each $15 {\rm \ km \ s^{-1}}$ channel.  Because the circumnuclear emission has a small angular extent compared with the primary beam, we did not apply any primary beam corrections to the maps.

\subsection{Radio observations: Submillimeter Array}
\label{sec_smaobs}
We observed NGC~7552 in $^{12}$CO~(2--1) and $^{13}$CO~(2--1) in the compact configuration of the Submillimeter Array  \cite[SMA;][]{Ho04} on 4 August 2006.  Observations were centered at $\rm \alpha_{2000}$=$23^{\rm h}16^{\rm m}10$\fs$7$ and $\rm \delta_{2000} = -42\arcdeg35\arcmin05\farcs41$, coinciding with the center of the circumnuclear starburst ring as defined by \citet{Fo94a}.  The primary beam of the antennas at the wavelengths of the observed molecular lines is 55\arcsec.  Seven out of the eight antennas of the SMA were available during our observation.   As summarized in Table \ref{TAB_obspara}, with baseline lengths ranging from 16~m to 69~m, the largest structure (of uniform brightness) that we expect to be sensitive to is about 16\arcsec.  The receiver was tuned and the correlator configured to cover the frequency range 227.6--229.6~GHz in the upper sideband (USB) thus including the $^{12}$CO~(2--1) line, and 217.6--219.6~GHz in the lower sideband (LSB) thus including the $^{13}$CO~(2--1) line.  The correlator provided a total of 24 spectral windows (chunks) in each sideband, with a total of 128 channels and a bandwidth of 104~MHz in each chunk.  To make the channel maps in $^{12}$CO~(2--1), we averaged 10 channels together resulting in a velocity resolution of $10 {\rm \ km \ s^{-1}}$.  Because the signal is weaker in $^{13}$CO~(2--1), to make the channel maps in this line we averaged 18 channels together resulting in a velocity resolution of $18 {\rm \ km \ s^{-1}}$.

We calibrated the data separately for bandpass, amplitude, and phase using SMA-specific MIR tasks adopted from the MMA software package (written in IDL), which was originally developed for the Owens Valley Radio Observatory \citep[OVRO,][]{Sco93}. The bandpass was calibrated using Uranus.  For complex gain calibration, we observed the quasars J2235--485 and J2258--279 for a duration of 6~min every 30~min.  We used J2235--485, which is located at an angular distance of 9$\fdg$3 from our target source, for phase calibration.  J2235--485, however, is too weak for amplitude calibration.  We therefore used J2258--279, which is located at a larger angular distance of 15$\arcdeg$ from our target source, for amplitude calibration.  Because the amplitude solutions derived from the individual scans did not differ significantly between adjacent scans, we applied time smoothing scale of three scans of J2258--279 to derive more precise amplitude solutions.   We derived the absolute flux calibration scale from Uranus, which we assumed had a flux density of 36.836~Jy at the time of our observation.

We imaged the data using MIRIAD. To deconvolve the sidelobes, we first CLEANed the dirty maps without any preselection to judge which features are likely to be real and which are likely to be sidelobes.  We then placed boxes around those features judged to be real on the basis that they do not appear to resemble the sidelobes of the dirty beam, and that their structure changed continuously between adjacent channels.  As shown in $\S$\ref{subsec_CO_result}, the features in each channel have a sufficiently simple structure that we are confident of having picked out only those that are real.  The $^{12}$CO~(2--1) channel maps have an angular resolution of 7$\farcs$0 $\times$ 2$\farcs$8 at PA = $-11\fdg9$, and a rms noise level of 78 mJy beam$^{-1}$ at a velocity resolution of 10 km s$^{-1}$.   The $^{13}$CO~(2--1) channel maps were made with 18 km s$^{-1}$ velocity binning, resulting in an angular resolution of 6$\farcs$9 $\times$ 2$\farcs$8 at a PA = $-8\fdg9$, and a rms noise level of 39 mJy beam$^{-1}$.  Because the features detected span an angular extent from the pointing center that is significantly smaller than the primary beam, we did not apply a primary beam correction to the maps.

\section{Results} 
\label{sec_result}
\subsection{Optical images}
\label{subsec_opt_result}
\begin{figure}
\plottwo{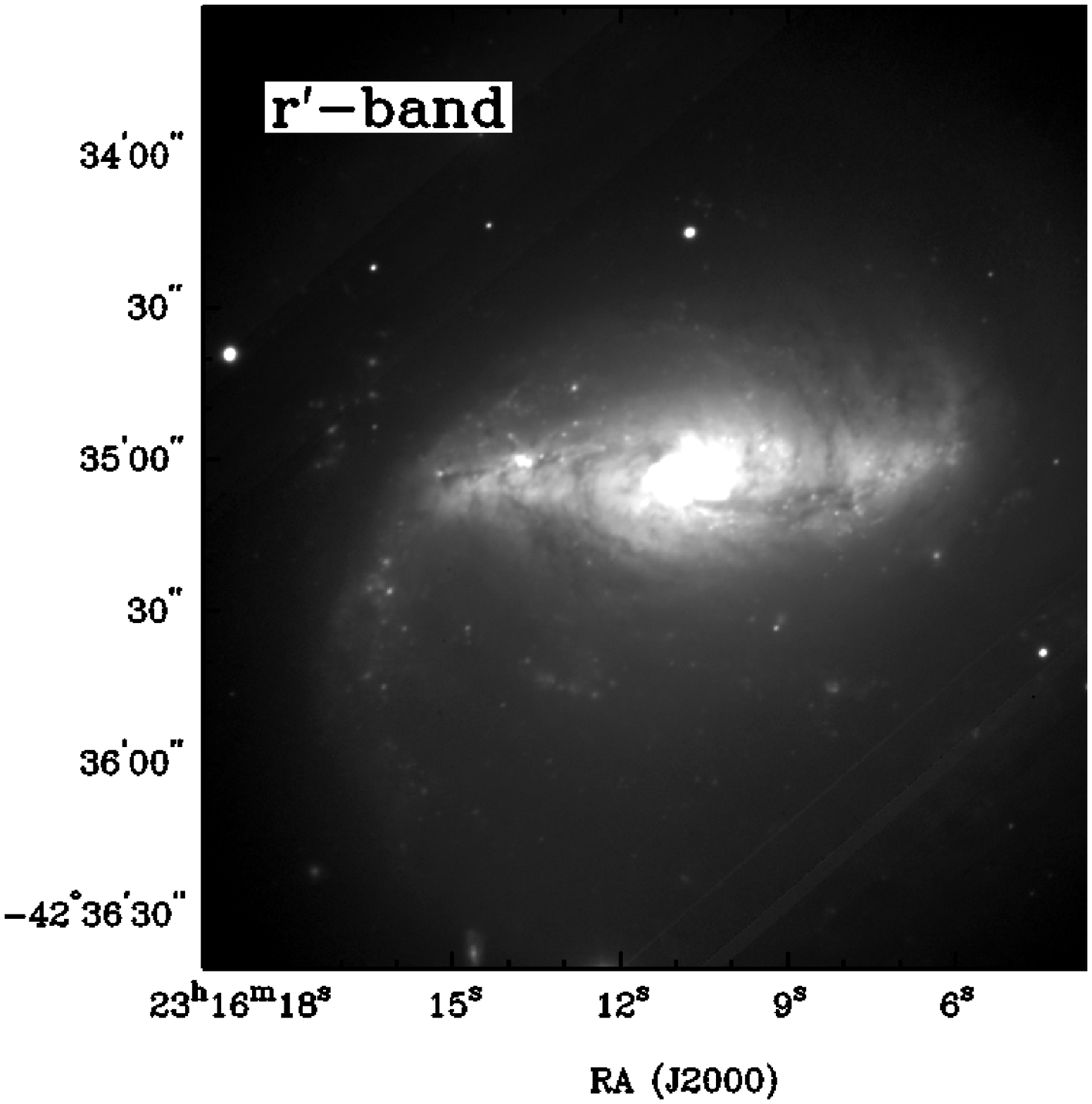}{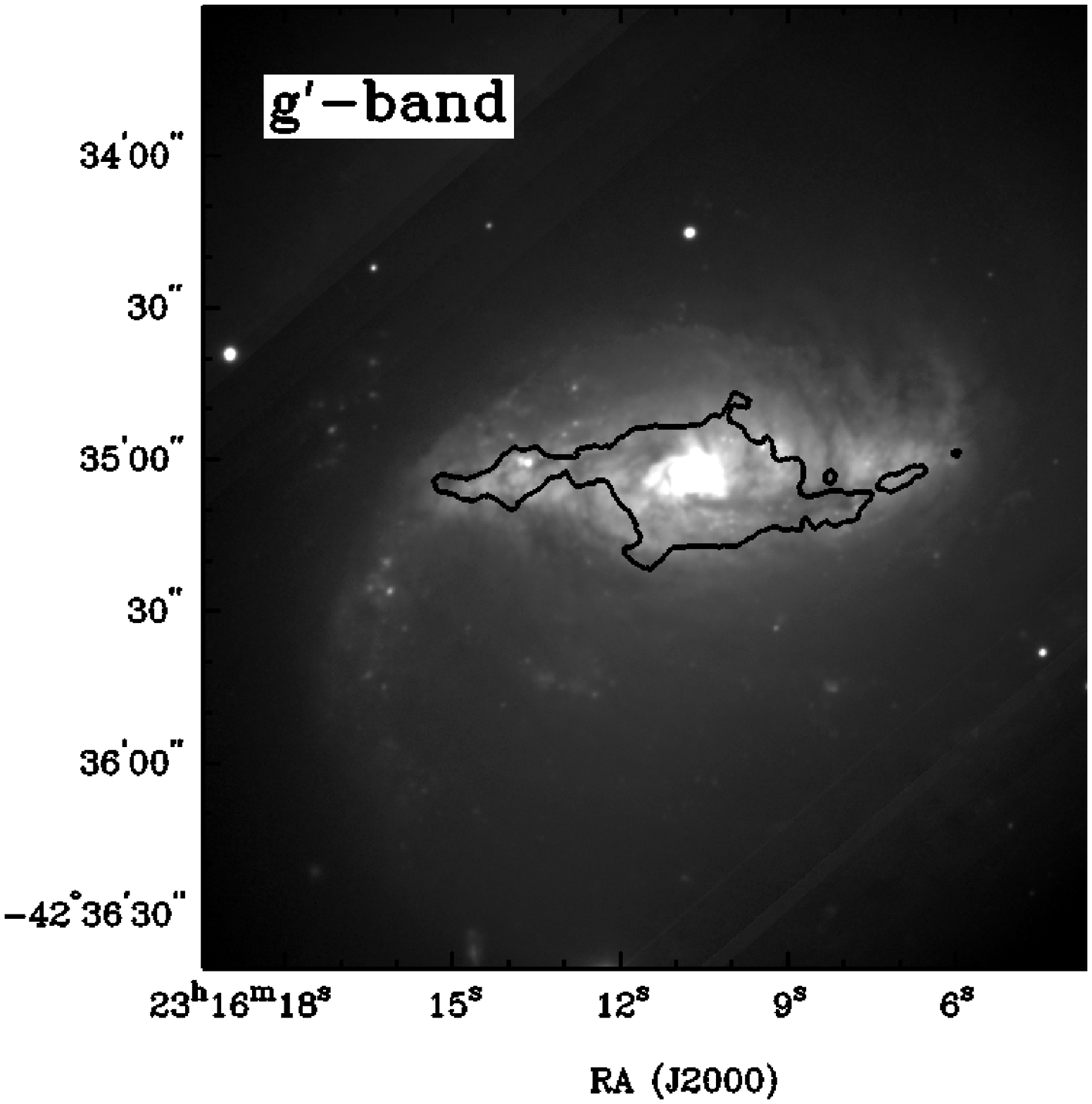}
\newline
\plottwo{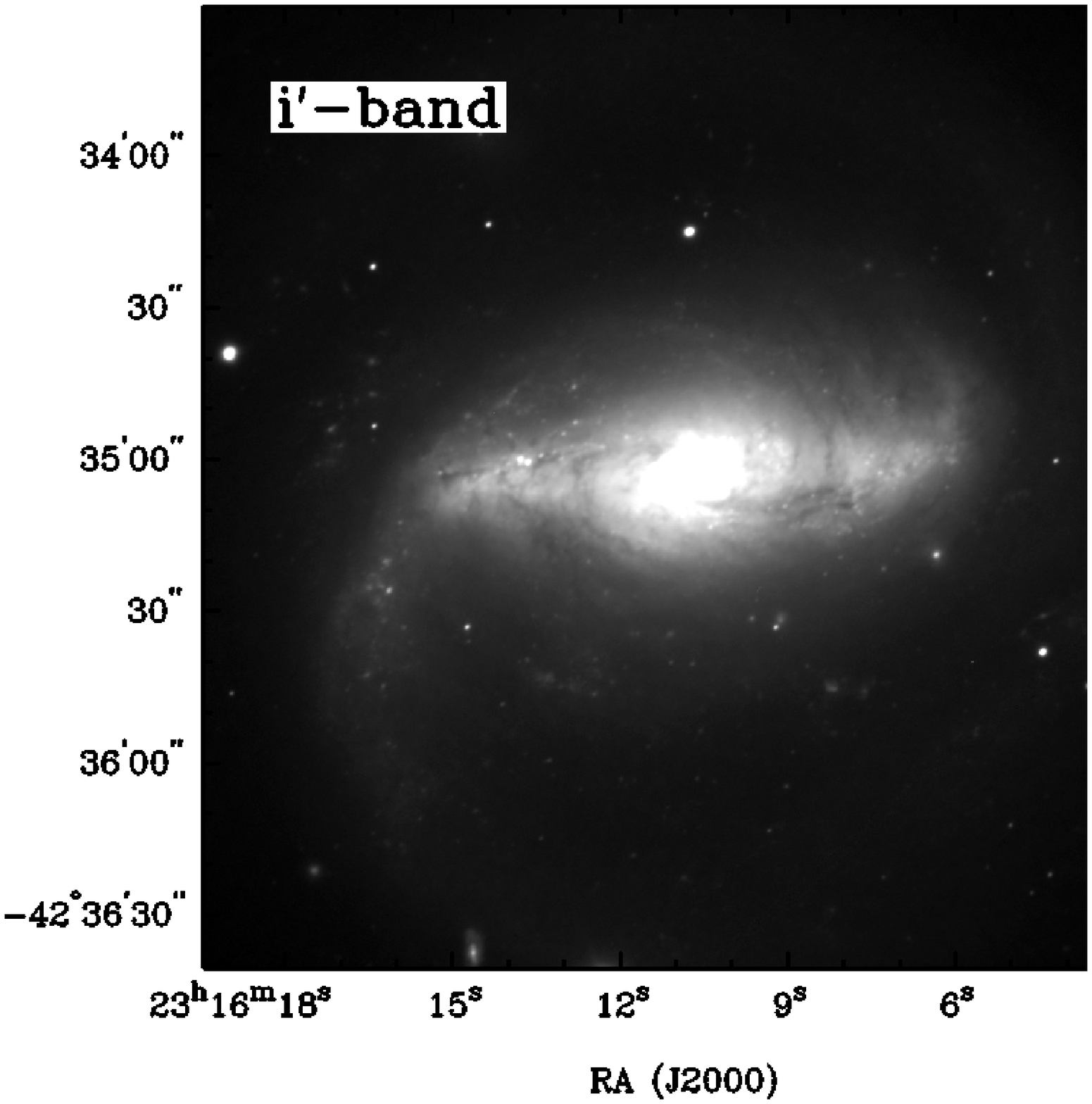}{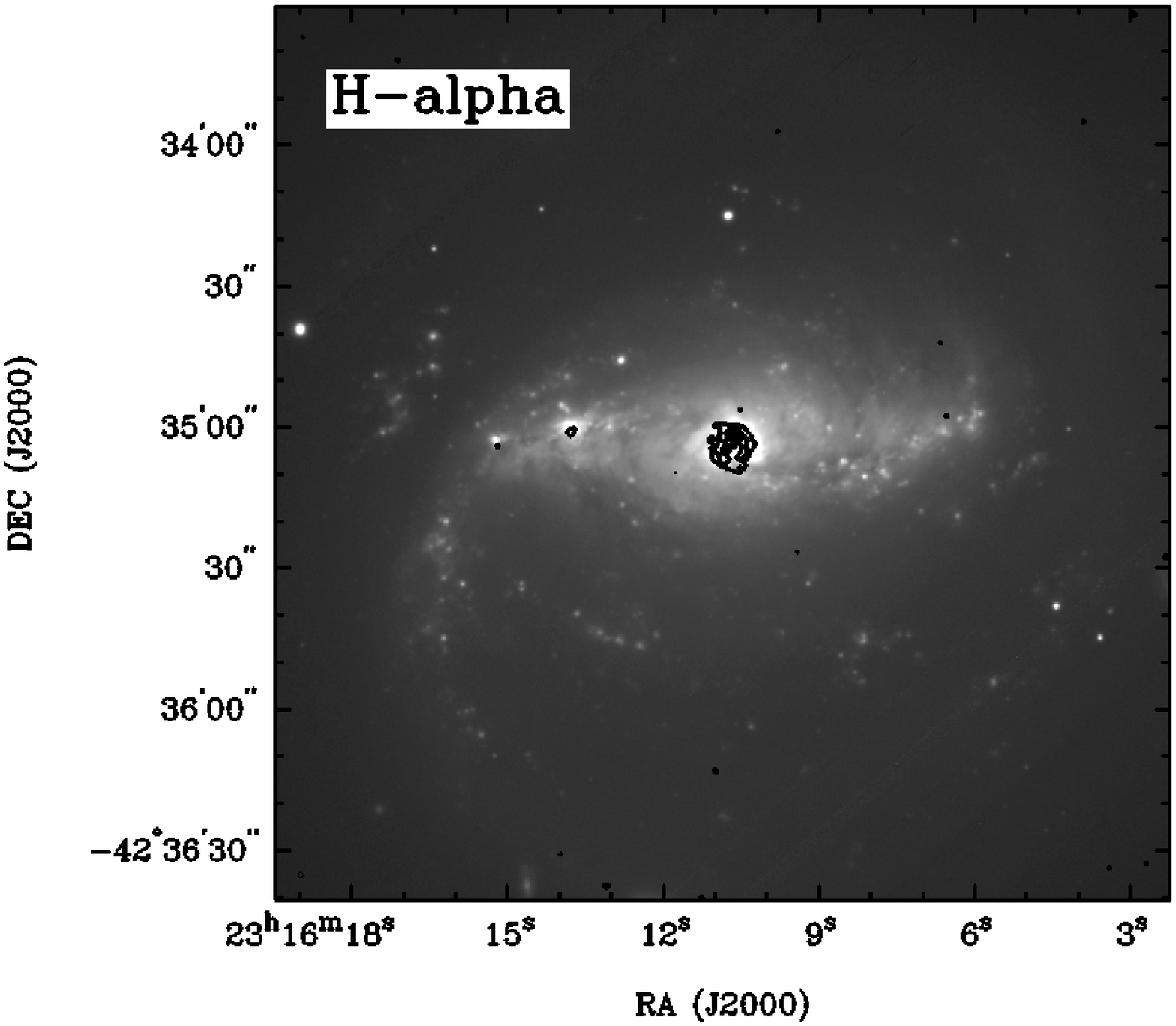}
\caption{Optical $r^{\prime}$- (upper left), $g^{\prime}$- (upper right), $i^{\prime}$-band (lower left) and H$\alpha$ (lower right) images of NGC 7552 taken with GMOS on the Gemini South telescope. The {\em Spitzer} 5.8 $\mu$m contour at 5.5 MJy sr$^{-1}$ \citep{Ken03} is overlaid on the $g^{\prime}$-band image, highlighting the orientation of the dust lanes. Likewise, the 3-cm continuum contours (\S\ref{subsec_continuum_result}; see text) showing the circumnuclear starburst ring are plotted in the H$\alpha$ image, with contour steps of 3, 15, and 35 $\sigma$, where 1 $\sigma$ = 66 $\mu$Jy beam$^{-1}$. \label{FIG_Optical}}
\end{figure}

Figure \ref{FIG_Optical} shows the optical images obtained with Gemini South.  Distinct dust lanes are often seen in late-type barred galaxies along the leading edges of the bar, as is the case also for NGC~7552.  In the g${}'$-band image of Figure \ref{FIG_Optical}, we indicate the location of the dust lanes with a more direct tracer, dust emission as imaged by $Spitzer$ at 5.8 $\mu$m.  The overall extent of this dust emission is indicated by a contour, which perfectly matches the dust lanes seen in the optical.

The H$\alpha$ image of Figure \ref{FIG_Optical} highlights regions of recent activity involving massive stars.  The circumnuclear starburst ring as traced in the 3-cm continuum is overlaid on the H$\alpha$ image to show the overall extent of this ring.  The physical size of the ring at 3 cm is comparable to the size of the central bright bulge of NGC~7552 in the optical images.   A magnified view of the central region in H$\alpha$ and 3-cm continuum is shown in Figure \ref{FIG_Continuum}.  In addition to the circumnuclear starburst ring, a bright \ion{H}{2} region, detected in both H$\alpha$ and radio continuum, is apparent $\sim$3.5~kpc east of the nucleus along the bar.  This \ion{H}{2} region was also detected in earlier H$\alpha$ images by \citet{Fo94a}.
\subsection{\ion{H}{1}}
\label{subsec_HI_result}
\begin{figure}
\plottwo{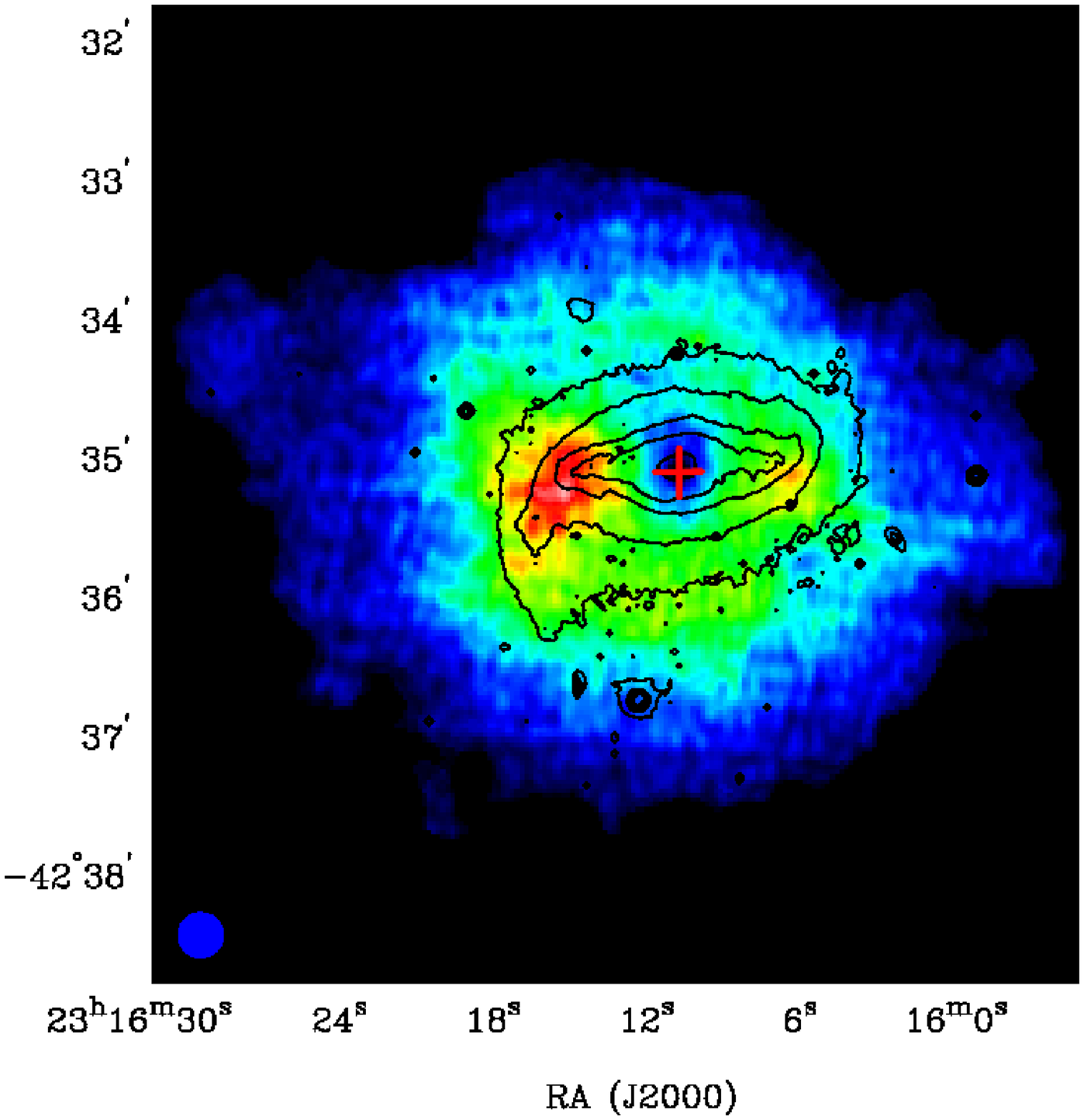}{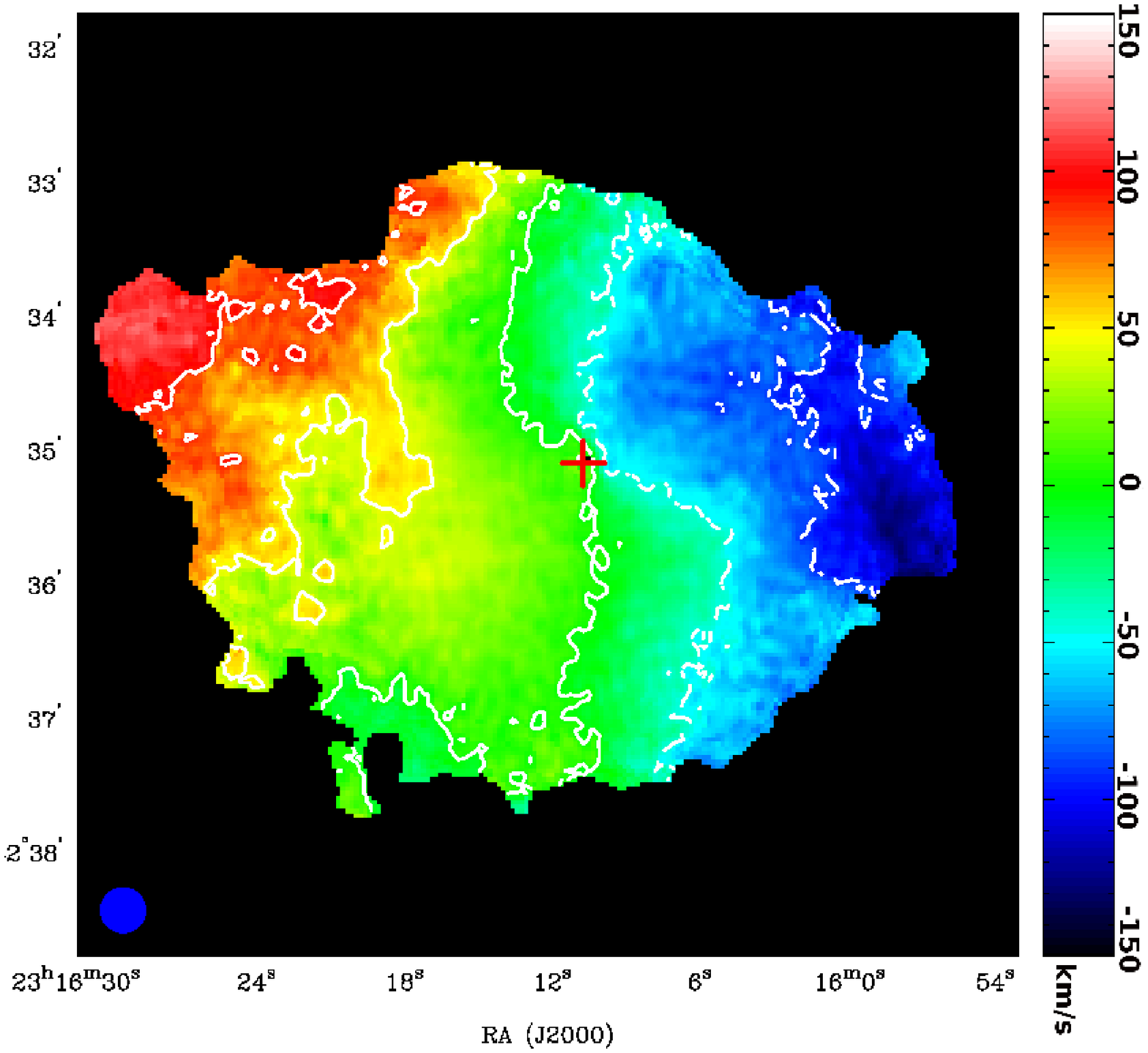}
\caption{Integrated \ion{H}{1} intensity (left panel) and intensity-weighted mean \ion{H}{1} velocity (right panel), both in color.
In the left panel, contours are of the $i^{\prime}$ image shown in Figure \ref{FIG_Optical}. 
In the right panel, contours correspond to the intensity-weighted mean \ion{H}{1} velocity, just like the color map, and are plotted in steps of -90, -45, 0, 45, and 90 km s$^{-1}$ with respect to the systemic velocity.  In each panel, a cross is plotted at the center of the galaxy, corresponding to the center of the starburst ring \citep{Fo94a}, and the synthesized beam shown at the lower left corner.\label{FIG_HI_MOM}}
\end{figure}

A map of the integrated \ion{H}{1} intensity (zeroth moment) is shown in Figure~\ref{FIG_HI_MOM} (left panel). As pointed out by \citet{Dah05}, the \ion{H}{1} gas traces the large-scale disk of NGC~7552.  For ease of comparison, the optical $i^{\prime}$-band image of the galaxy shown earlier in Figure~\ref{FIG_Optical} is overlaid as contours.  Along the major axis the \ion{H}{1} disk can be detected out to a projected radius of about 2100\arcsec\ ($\sim$20~kpc), about  twice as far out as the optical disk.  Similarly, along the minor axis, the \ion{H}{1} disk can be detected out to a projected radius of about 121\arcsec\ ($\sim$13~kpc), again about twice as far out as the optical disk.  There is a prominent ring-like enhancement in \ion{H}{1} 
that peaks at or close to the ends of the large-scale optical bar.  Finally, a central depression can be seen in the integrated \ion{H}{1} intensity map.  This central depression has a size that closely coincides with the angular extent of the circumnuclear molecular gas detected in our observations as described below ($\S$\ref{subsec_hcn_result} and $\S$\ref{subsec_CO_result}).  

Figure~\ref{FIG_HI_MOM} (right panel) shows a map of the intensity-weighted mean \ion{H}{1} velocity (first moment). Assuming a systemic velocity of $1586 {\rm \ km \ s^{-1}}$ as found in the optical and as derived below ($\S$\ref{subsec_CO_result}) from our molecular line observations, emission can be seen spanning the velocity range from $-100 {\rm \ km \ s^{-1}}$ to $100 {\rm \ km \ s^{-1}}$ about the systemic velocity. As can be seen, the kinematic major axis is aligned roughly east-west, coincident with the major axis of the optical bar. Blueshifted emission (negative velocities) originates primarily westwards and redshifted emission (positive velocities) primarily eastwards of the galactic center. On close scrutiny, the position angle of the kinematic major axis can be seen to change with galactic radius, creating an S-shaped pattern in the gas kinematics.  Such S-shaped kinematic patterns are commonly attributed to non-circular motions under the non-axisymmetric gravitational potential of a galactic bar (see $\S$\ref{subsec_brandt_rotation_curve} for detailed discussion).  Far beyond the optical bar however, the position angle of the kinematic major axis continues to change, suggesting that the outskirts of the galaxy may be warped (by gravitational interactions with one or more neighboring galaxies as described in $\S$\ref{sec_intro}).

\subsection{Radio Continuum}
\label{subsec_continuum_result}
\begin{figure}
\plottwo{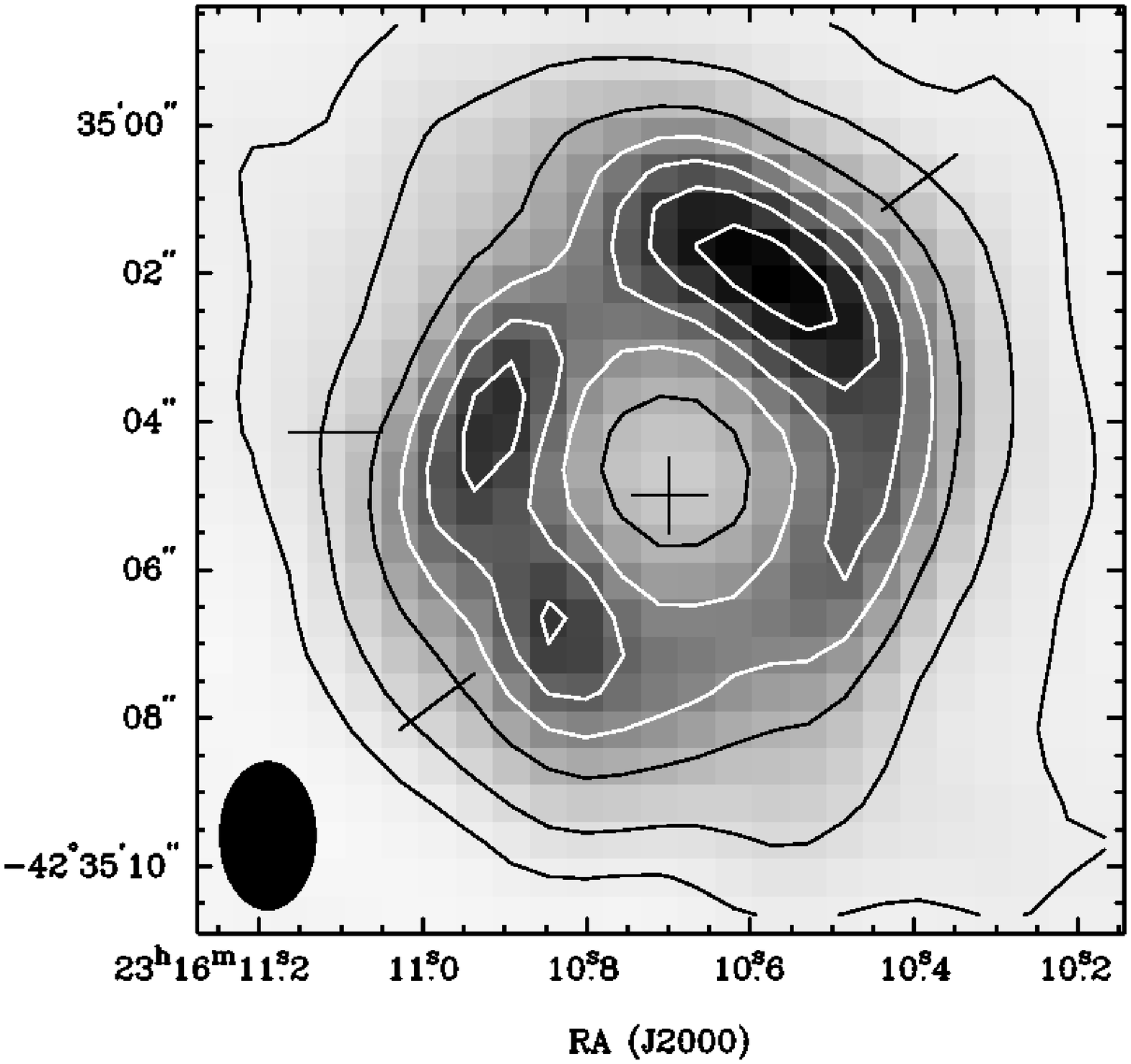}{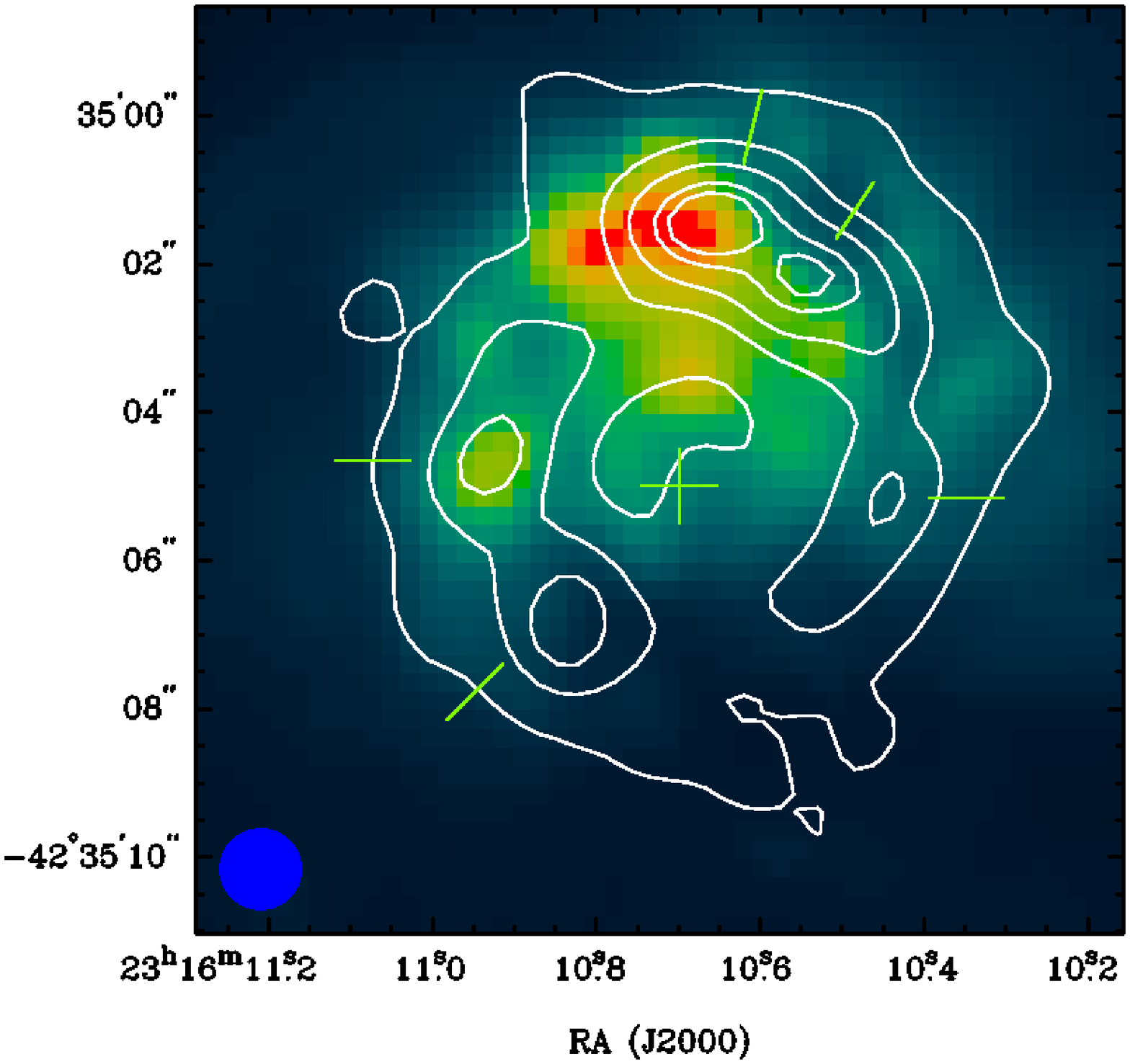}
\caption{Radio continuum images of the circumnuclear starburst ring in NGC~7552 at 6~cm (left panel; both contours and greyscale) and 3~cm (right panel; contours) recreated from the data taken by and published in \citet{Fo94a}.  In the right panel, the 3-cm (contours) is overlaid on the H$\alpha$ (color scale) image shown in Figure~\ref{FIG_Optical}.  Straight lines in both panels indicate peaks in radio emission, which correspond to star-forming knots.
Contour levels of the 6-cm image are at 3, 10, 25, 40, 55, 70, and 85$\sigma$, where 1 $\sigma$ = 72 $\mu$Jy beam$^{-1}$, and contour levels of the 3-cm image are at 3, 5, 15, 25, 35, and 40 $\sigma$, where 1 $\sigma$ = 66 $\mu$Jy beam$^{-1}$.
A cross in each panel indicates the center of the ring as defined by \citet{Fo94a}.  The synthesized beam is $2\farcs0 \times 1\farcs3$ at 6~cm  and $1\farcs1 \times 1\farcs1$ at 3~cm, and shown at the lower left corner of the respective panels.\label{FIG_Continuum}}
\end{figure}

Figure~\ref{FIG_Continuum} shows the central part of our radio continuum images at 6~cm (left panel) and 3~cm (right panel), both of which were recreated from the data taken by \citet{Fo94a}.  Over the field shown, the radio continuum emission originates from a ring, which as mentioned by \citet{Fo94a} corresponds to the circumnuclear starburst ring, as can be seen in Figure~\ref{FIG_Continuum} (right panel) where the 3-cm image (contours) is overlaid on our H$\alpha$ image (color).
This ring has an outer extent comparable to the central \ion{H}{1} depression mentioned above.  

Five peaks are clearly visible in the 3-cm image (right panel), whereas only three peaks are visible in the 6-cm image (left panel) that has a lower angular resolution than the 3-cm image.  To study the spectral index of each knot, we have convolved the 3-cm image to the same angular resolution as the 6-cm image.  After degrading the angular resolution, we find that only three knots are discernible in the-3 cm image, which now resembles the 6-cm image. The spectral indices ($\alpha$, where $S\propto \nu^{\alpha }$ with $S$ the flux density and $\nu$ the observing frequency) of the three knots (see the left panel of Fig.~\ref{FIG_Continuum}) are $-1.11 \pm 0.02$ (upper right), $-0.84 \pm 0.03$ (upper left), and $-0.72 \pm 0.05$ (bottom left), with an average value of $-0.89 \pm 0.06$ that is similar to that inferred by \citet{Fo94a}.  The negative spectral indices indicate that the emission of all the knots is dominated by, if not produced entirely by, (non-thermal) synchrotron emission.
\subsection{HCN and HCO$^{+}$}
\label{subsec_hcn_result}
\begin{figure}
\plotone{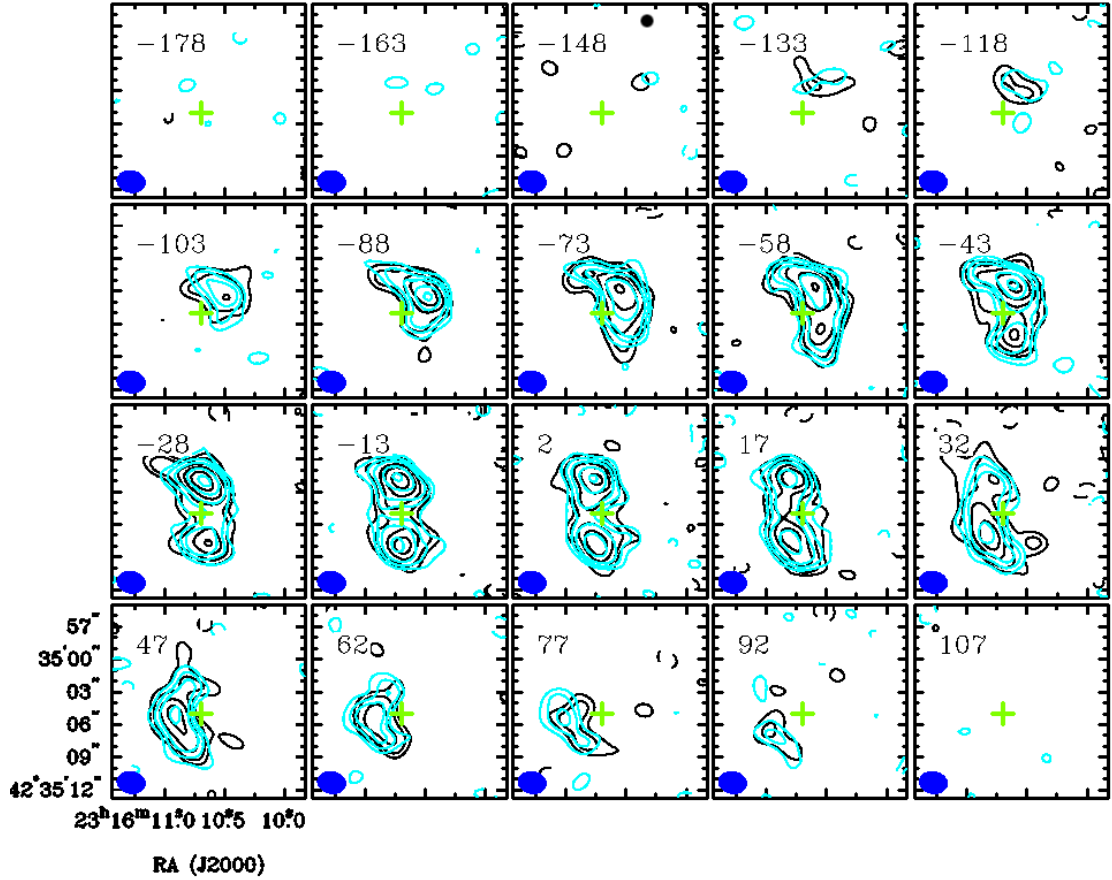}
\caption{Channel maps of the central region of NGC~7552 in HCN~(1 -- 0) (black contours) and HCO$^{+}$~(1 -- 0) (cyan contours).  Contour levels are plotted at -2, 2, 3, 5, 7, and $10 \times 4.2 {\rm \  mJy \ beam^{-1}}$ (rms noise level) for HCN~(1 -- 0) and $\times 4.5 {\rm \  mJy \ beam^{-1}}$ (rms noise level) for HCO$^{+}$~(1 -- 0).  The velocity of each channel, in units of km s$^{-1}$, is shown in the upper left corner of each panel.    A cross is plotted at the center of the galaxy in each panel.  The synthesized beam is $2\farcs6 \times 2\farcs0$ at a position angle of 80\degr, and is shown at the lower left corner of each panel.\label{FIG_HCN.CM}}
\end{figure}

Figure~\ref{FIG_HCN.CM} shows our channel maps of the HCN (black contours) and HCO$^{+}$~(1 -- 0) (cyan contours) lines.  Both HCN~(1 -- 0) and HCO$^{+}$~(1 -- 0) emission are concentrated in two knots located to the north and south of the galactic nucleus.  The northern knot is detectable over a velocity range from $-133 {\rm \ km \ s^{-1}}$ to $32 {\rm \ km \ s^{-1}}$ with respect to the optically-determined systemic velocity of the galaxy, and is therefore preferentially blueshifted.  The southern knot is detectable over a velocity range from about $-43 {\rm \ km \ s^{-1}}$ to $92 {\rm \ km \ s^{-1}}$ with respect to the systemic velocity, and is therefore preferentially redshifted.  The centroids of both knots shift from west to east with increasing (i.e., from blueshifted to redshifted) velocities. 

\begin{figure}
\plottwo{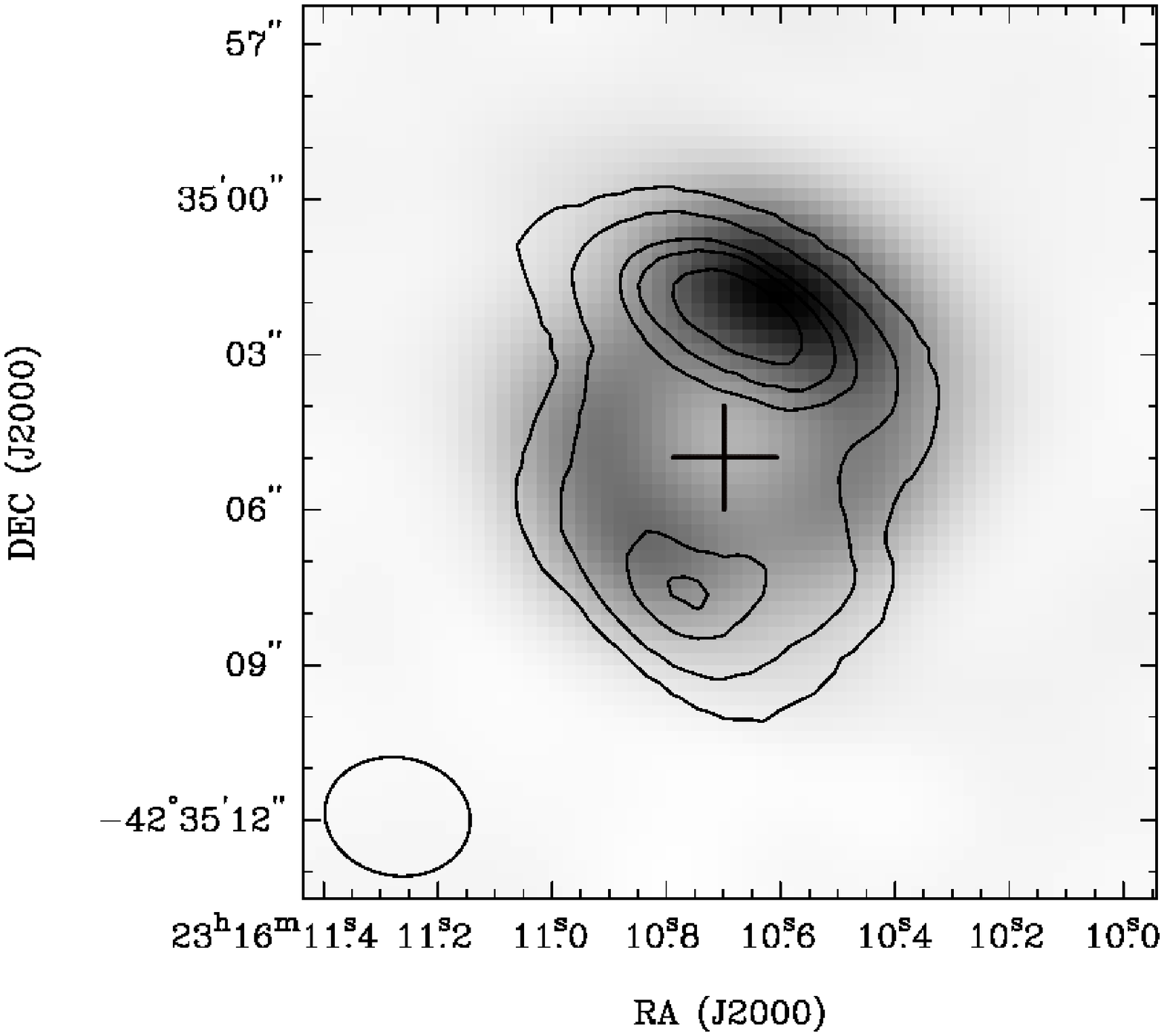}{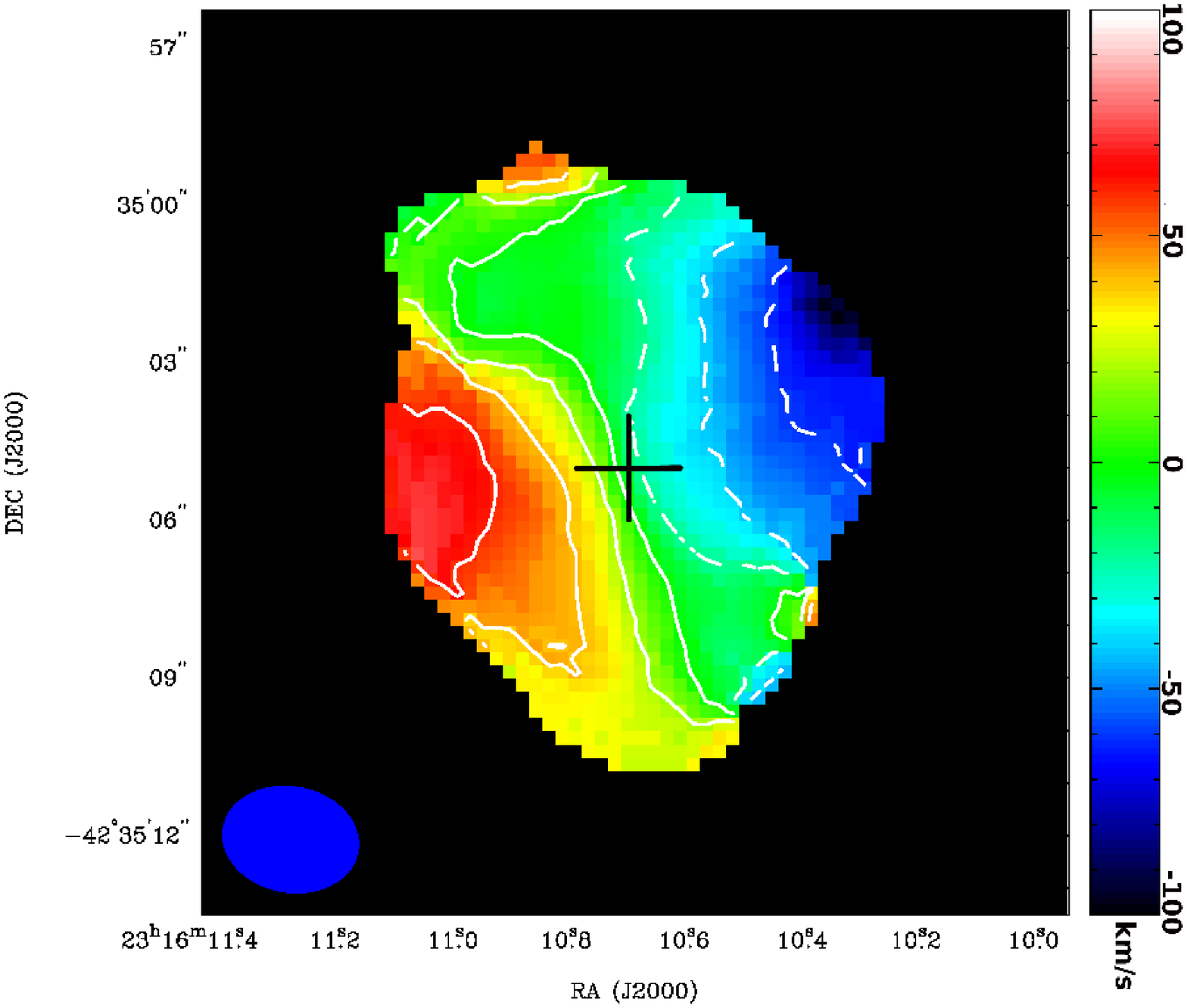}
\caption{Integrated HCN~(1 -- 0) intensity (left panel; contours) and intensity-weighted mean HCN~(1 -- 0) velocity (right panel; contours and color scale) derived from the channel maps shown in Figure~\ref{FIG_HCN.CM}. 
In the left panel, contours indicating the integrated HCN~(1 -- 0) intensity are plotted at 3, 6, 10, 12, and 15 $\sigma$, where 1 $\sigma$ = 0.3 Jy beam$^{-1}$ km s$^{-1}$.  Greyscale corresponds to the 3-cm continuum image shown in Figure~\ref{FIG_Continuum}, but convolved to the same angular resolution as the HCN~(1 -- 0) image.   In the right panel, contours correspond to the intensity-weighted mean HCN~(1 -- 0) velocity, just like the color map, and are plotted in steps of $20 {\rm \ km \ s^{-1}}$ starting from $-60 {\rm \ km \ s^{-1}}$ and ending at $60 {\rm \ km \ s^{-1}}$ with respect to the systemic velocity.   A cross is plotted at the center of the galaxy, and the synthesized beam shown at the lower left corner, of each panel. \label{FIG_HCN_MOM}}
\end{figure}

Figure~\ref{FIG_HCN_MOM} (left panel) shows a map of the integrated HCN~(1 -- 0) intensity in black contours.  The overall size of the emitting region is comparable with the central \ion{H}{1} depression described above in $\S$\ref{subsec_HI_result}, suggesting that the gas density is so high here that the atomic hydrogen gas has been largely converted to molecular hydrogen gas.  The  HCN~(1 -- 0) contours is overlaid on a greyscale image of the 3-cm continuum emission shown earlier in Figure~\ref{FIG_Continuum} (right panel) but now convolved to the same angular resolution as the HCN~(1 -- 0) map.  At this angular resolution, the 3-cm continuum map shows two knots just like the HCN~(1 -- 0) map.  The observed HCN~(1 -- 0) emission therefore originates from the circumnuclear starburst ring, and exhibits a central depression or possibly a hole.  Notice the small displacement between the centroids of the HCN~(1 -- 0) gas and radio continuum knots, a point we shall return to in $\S$\ref{sec_discussion_Formation}. 

Figure~\ref{FIG_HCN_MOM} (right panel) shows a map of the intensity-weighted mean HCN~(1 -- 0) velocity.  A velocity gradient is discernible roughly along the east-west direction, with the kinematic major axis lying at a position angle of 110$\arcdeg$. The kinematic major axis in HCN~(1 -- 0), as well as in HCO$^{+}$~(1 -- 0) as described below, are both aligned with the kinematic major axis at the innermost region of the large-scale \ion{H}{1 } disk as described in $\S$\ref{subsec_HI_result}.  The good agreement suggests that there is a continuous velocity field between the outermost region in molecular gas and innermost region in neutral atomic gas.

\begin{figure}
\plottwo{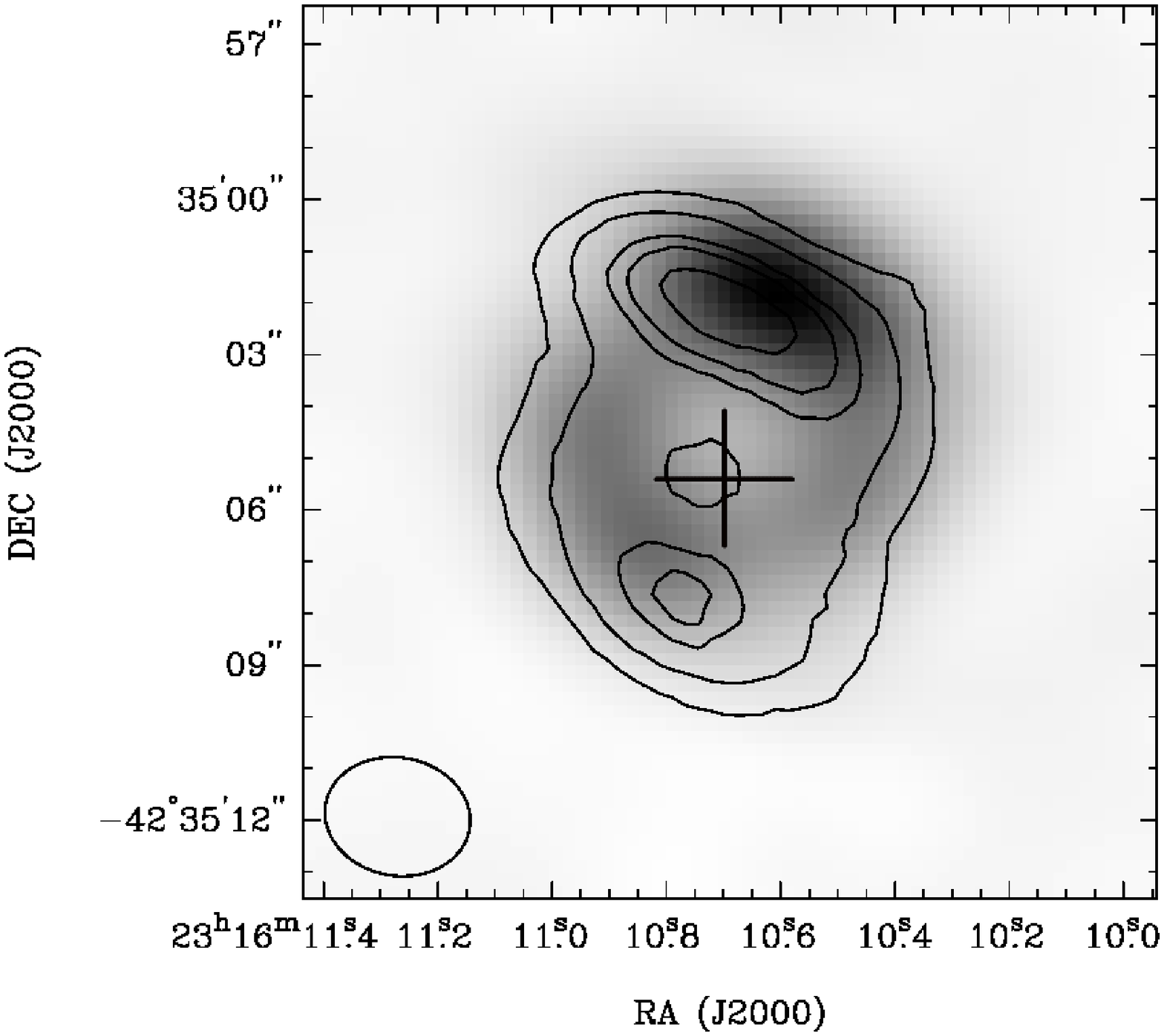}{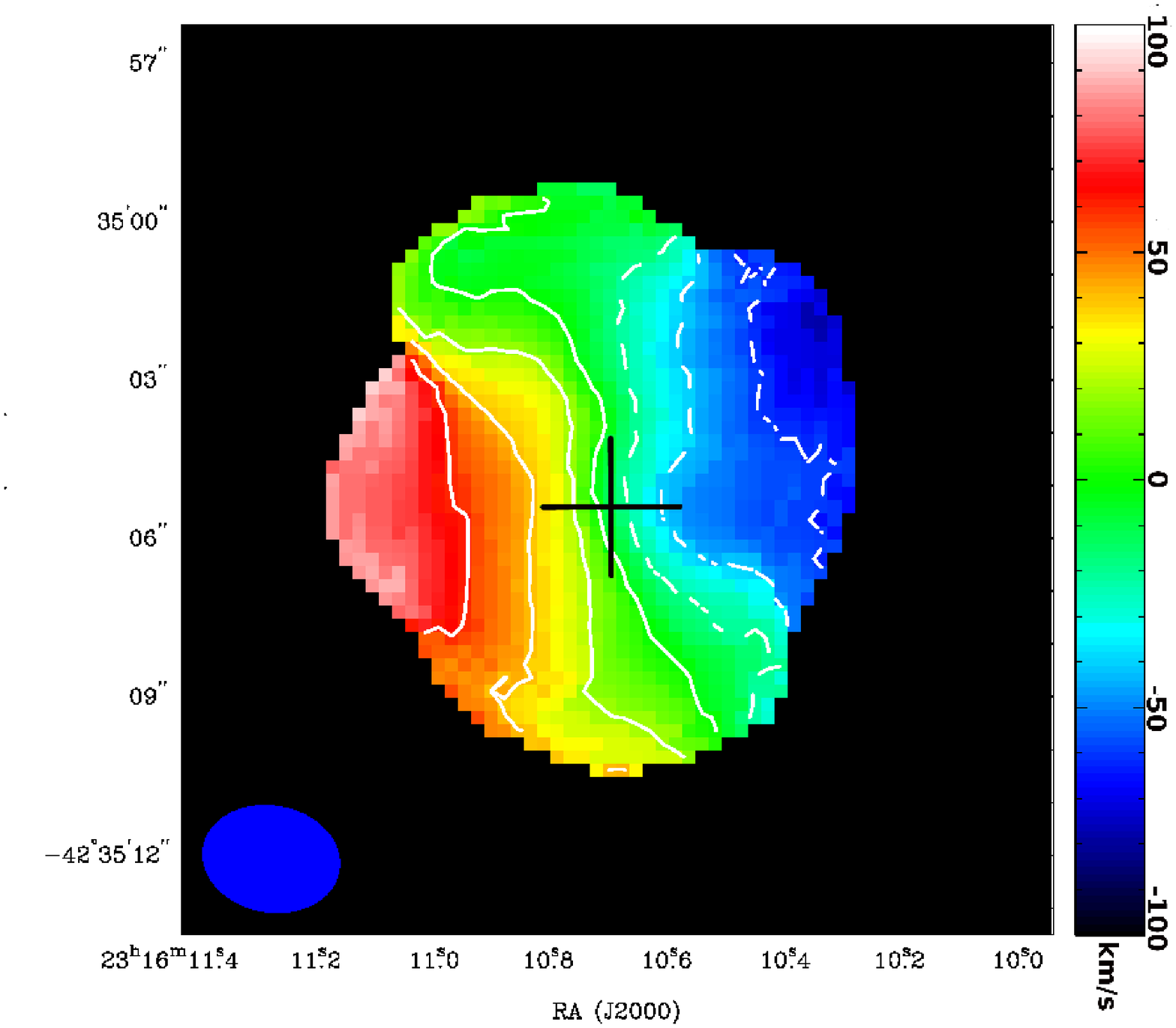}
\caption{ntegrated HCO$^+$~(1 -- 0) intensity (left panel; contours) and intensity-weighted mean HCO$^+$~(1 -- 0) velocity (right panel; contours and color scale) derived from the channel maps shown in Figure~\ref{FIG_HCN.CM}.   In the left panel, contours indicating the integrated HCO$^+$~(1 -- 0) intensity are plotted at 3, 6, 10, 12, and 15 $\sigma$, where $1 \sigma = 0.3$~Jy~beam$^{-1}$~km~s$^{-1}$.
Greyscale corresponds to the 3-cm continuum image shown in Figure~\ref{FIG_Continuum}, but convolved to the same angular resolution as the HCO$^+$~(1 -- 0) image.  
In the right panel, contours correspond to the intensity-weighted mean HCO$^+$~(1 -- 0) velocity, just like the color map, and are plotted in steps of $20 {\rm \ km \ s^{-1}}$ starting from $-60 {\rm \ km \ s^{-1}}$ and ending at $60 {\rm \ km \ s^{-1}}$ with respect to the systemic velocity.  A cross is plotted at the center of the galaxy, and the synthesized beam shown at the lower left corner, of each panel.\label{FIG_HCO_MOM}}
\end{figure}

Figure~\ref{FIG_HCO_MOM} shows our moment maps of the HCO$^{+}$~(1 -- 0) line. The emission in all these maps is virtually identical to that seen in HCN~(1 -- 0).  Both HCN~(1 -- 0) and HCO$^{+}$~(1 -- 0) have comparable critical densities of $\sim$$10^6 \rm \ cm^{-3}$, suggesting that both are tracing the same regions of dense gas.
\subsection{CO}
\label{subsec_CO_result}
\begin{figure}
\plotone{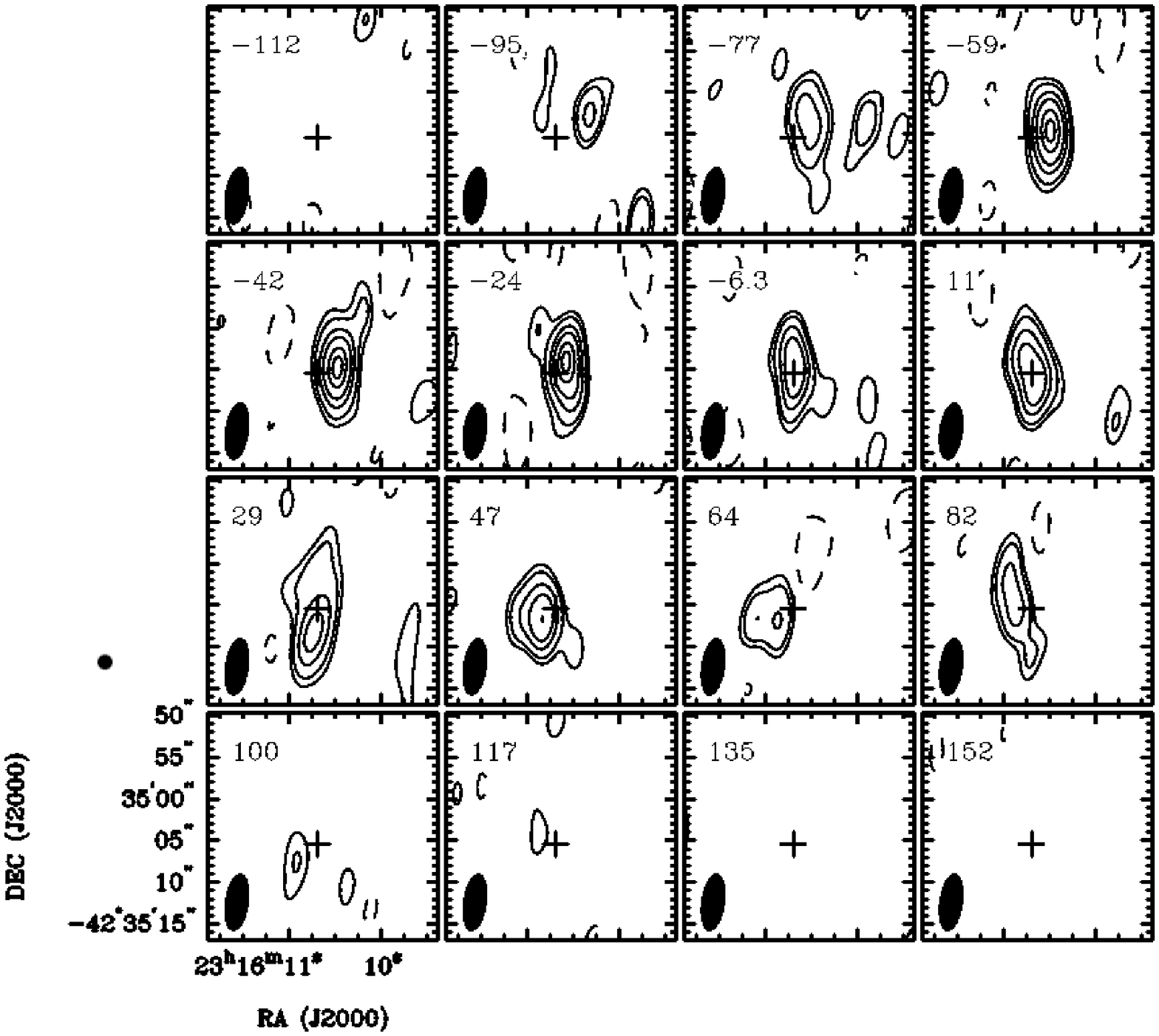}
\caption{Channel maps of the central region of NGC~7552 in $^{13}$CO~(2 -- 1).  Contour levels are plotted at --2, 2, 3, 5, 7, 10, and $12 \times 39 {\rm \  mJy \ beam^{-1}}$ (rms noise level).  The velocity of each channel, in units of km s$^{-1}$, is shown in the upper left corner of each panel.    A cross is plotted at the center of the galaxy in each panel.   The synthesized beam is $6\farcs9 \times 2\farcs8$ at a position angle of --8\fdg1, and is shown at the lower left corner of each panel.\label{FIG_13CO.CM}}
\end{figure}

Figure~\ref{FIG_13CO.CM} shows our channel maps of the $^{13}$CO~(2 -- 1) line, which by contrast with both the HCN~(1 -- 0) and HCO$^{+}$~(1 -- 0) has a much lower critical density of just $\sim$$10^3 \rm \ cm^{-3}$.  Only a single compact component is discernible that shifts from west to east across the center of the galaxy with increasing velocity, just like that seen in HCN~(1 -- 0) and HCO$^{+}$~(1 -- 0) as described above in $\S$\ref{subsec_hcn_result}. We have convolved both the HCN(1--0) and HCO$^{+}$(1--0) maps to the same angular resolution as the $^{13}$CO~(2 -- 1) map, and found that all show the same spatial-kinematic structure suggesting that they all trace the same global features.  In the $^{13}$CO~(2 -- 1) map, emission is detectable over the velocity range from about  $-95 {\rm \ km \ s^{-1}}$ to $100 {\rm \ km \ s^{-1}}$ with respect to the systemic velocity, comparable to the detectable velocity range of the emission in both the HCN~(1 -- 0) and HCO$^+$~(1 -- 0) maps.

\begin{figure}
\plottwo{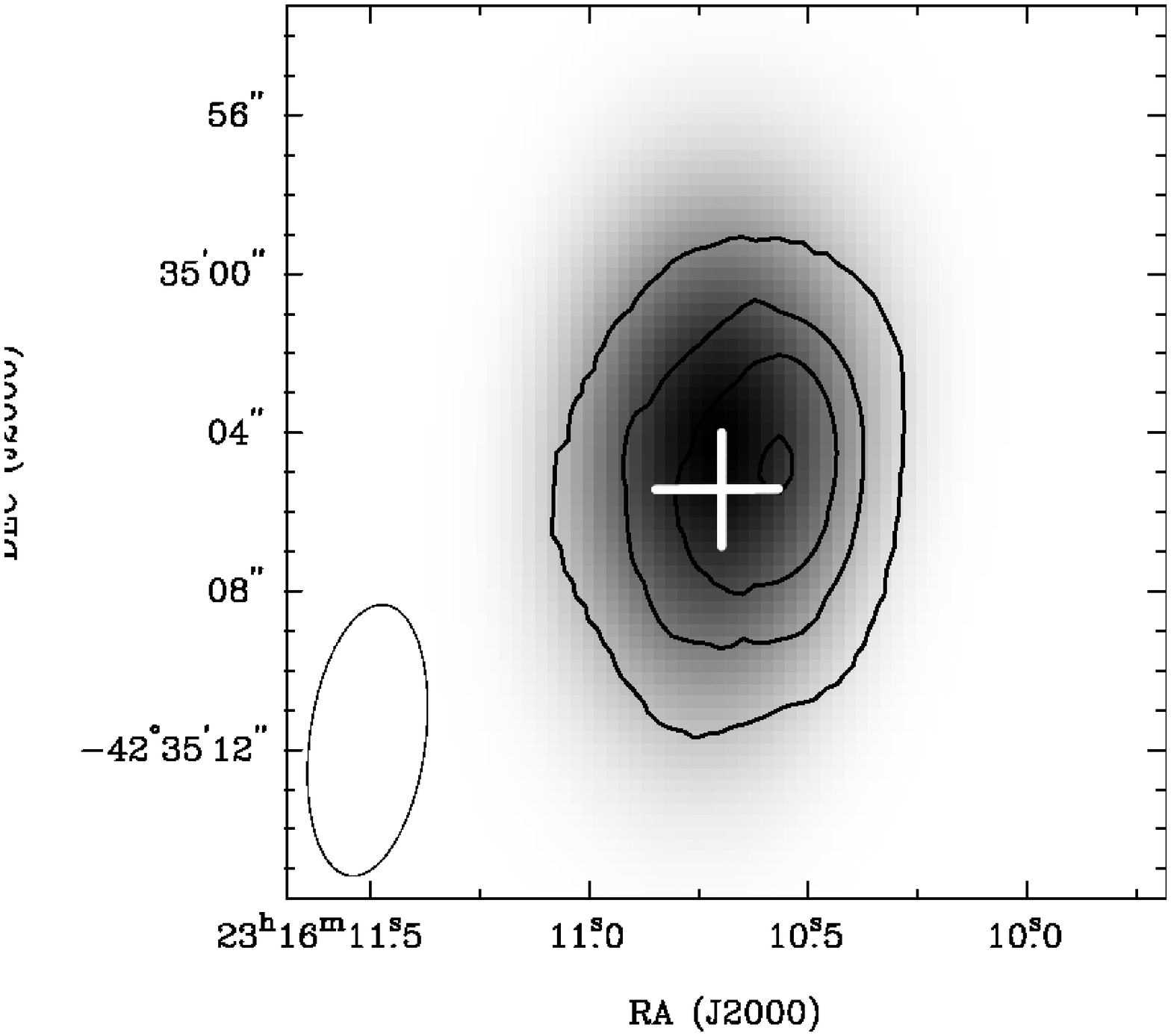}{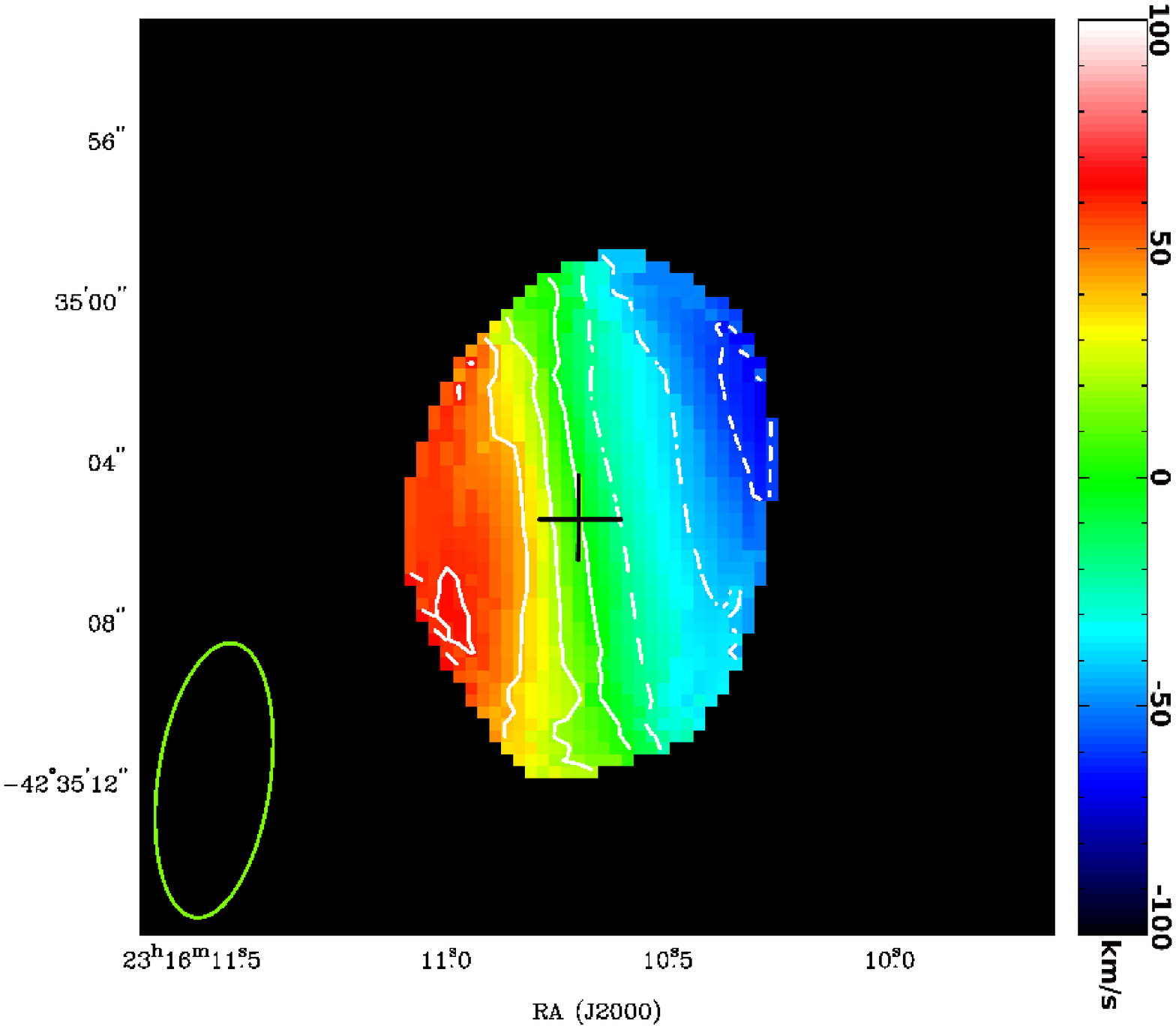}
\caption{Integrated $^{13}$CO~(2 -- 1) intensity (left panel; contours) and intensity-weighted mean $^{13}$CO~(2 -- 1) velocity (right panel; contours and color scale) derived from the channel maps shown in Figure~\ref{FIG_13CO.CM}.  In the left panel, contours  indicating the integrated $^{13}$CO~(2 -- 1) intensity are plotted at 3, 10, 14, and $18 \sigma$, where $1 \sigma = 2.3$~Jy beam$^{-1}$ km s$^{-1}$. 
Grayscale corresponds to the integrated HCN~(1 -- 0) intensity image shown in Figure~\ref{FIG_HCN_MOM}, but convolved to the same angular resolution as the $^{13}$CO~(2 -- 1) image.
In the right panel, contours correspond to the intensity-weighted mean $^{13}$CO~(2 -- 1) velocity, just like the color map, and are plotted in steps of $20 {\rm \ km \ s^{-1}}$ starting from $-60 {\rm \ km \ s^{-1}}$ and ending at $60 {\rm \ km \ s^{-1}}$ with respect to the systemic velocity.  A cross is plotted at the center of the galaxy, and the synthesized beam shown at the lower left corner, of each panel
\label{FIG_13CO_MOM}}
\end{figure}

\begin{figure}
\centering 
\includegraphics[scale=0.8]{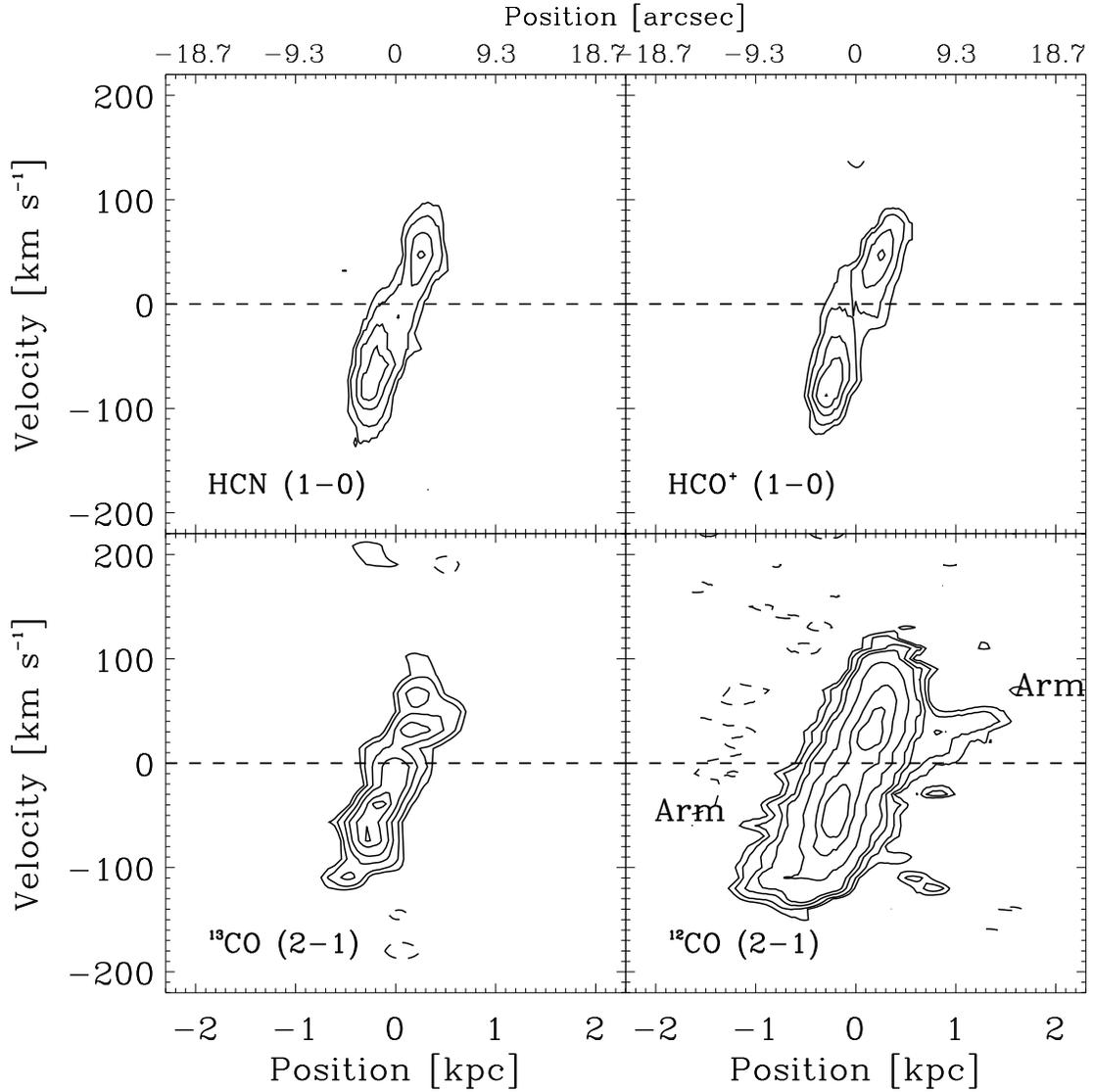}
\caption{
Position-velocity (P-V) diagrams of HCN~(1 -- 0) (upper left panel), HCO$^+$~(1 -- 0) (upper right panel), $^{13}$CO~(2 -- 1) (lower left panel) and $^{12}$CO~(2 -- 1) (lower right panel) along a position angle of $110\degr$, which corresponds to the orientation of the kinematic major axis.  Negative positions are located eastwards and positive positions located westwards of the galaxy center.   Dashed horizontal lines indicate 0 km s$^{-1}$ relative to the systemic velocity of 1580 km s$^{-1}$.  Contour levels of HCN (1 -- 0) and HCO$^+$~(1 -- 0) are at $-2$, 2, 3, 5, 7, and 10 $\sigma$, where 1 $\sigma$ is 4.2 and 4.5 mJy beam$^{-1}$, respectively, for HCN (1 -- 0) and HCO$^+$~(1 -- 0).  Contours levels of $^{13}$CO~(2 -- 1) are at --2, 2, 3, 5, 7, 10, 12, and 15 $\sigma$, where 1 $\sigma$ = 39 mJy beam$^{-1}$, and $^{12}$CO~(2 -- 1) at --2, 2, 3, 5, 7, 15, 30, and 40 $\sigma$, where 1 $\sigma$ = 78 mJy beam$^{-1}$. 
}
\label{FIG_PV_4lines}
\end{figure}

Figure~\ref{FIG_13CO_MOM} (left panel) shows the integrated $^{13}$CO~(2 -- 1) intensity map in black contours overlaid on the integrated HCN~(1 -- 0) intensity map, taken from Figure~\ref{FIG_HCN_MOM} but convolved to the same angular resolution as the $^{13}$CO~(2 -- 1) map,  in grayscale.  Figure~\ref{FIG_13CO_MOM} (right panel) shows a map of the intensity-weighted mean $^{13}$CO~(2 -- 1) velocity.  As can be seen, the compact central component detected in $^{13}$CO~(2 -- 1) has a similar size and kinematics as the circumnuclear HCN~(1 -- 0) (and HCO$^{+}$~(1 -- 0)) emission. In Figure~\ref{FIG_PV_4lines}, we show position-velocity (P-V) diagrams of the HCN~(1 -- 0), HCO$^+$~(1 -- 0), and $^{13}$CO~(2 -- 1) emission along a position angle of $110\degr$, which corresponds to the orientation of the kinematic major axis.  As can be seen, the P-V diagram in $^{13}$CO~(2 -- 1) closely resembles that in the higher density tracers of molecular hydrogen gas.  We therefore believe that the observed $^{13}$CO~(2 -- 1) emission originates from essentially the same regions of the circumnuclear starburst ring as the observed HCN~(1 -- 0) or HCO$^{+}$~(1 -- 0) emission.  Of course, any central depression or hole of the same size as that seen in HCN~(1 -- 0) and HCO$^{+}$~(1 -- 0) would not be discernible at the angular resolution of our $^{13}$CO~(2 -- 1) observation.  

The P-V diagrams of Figure~\ref{FIG_PV_4lines} reveal that the systemic velocity at the center of the circumnuclear starburst ring, which presumably coincides with the dynamical center of the galaxy, is about $1580 {\rm \ km \ s^{-1}}$.  This is consistent with the optically-determined value of $1586 \pm 5 {\rm \ km \ s^{-1}}$ (RC3), but significantly smaller than that determined from the global \ion{H}{1} profile of $1608 \pm 5 {\rm \ km \ s^{-1}}$ \citep[The \ion{H}{1} Parkes All Sky Survey;][]{Bar01}.

\begin{figure}
\plotone{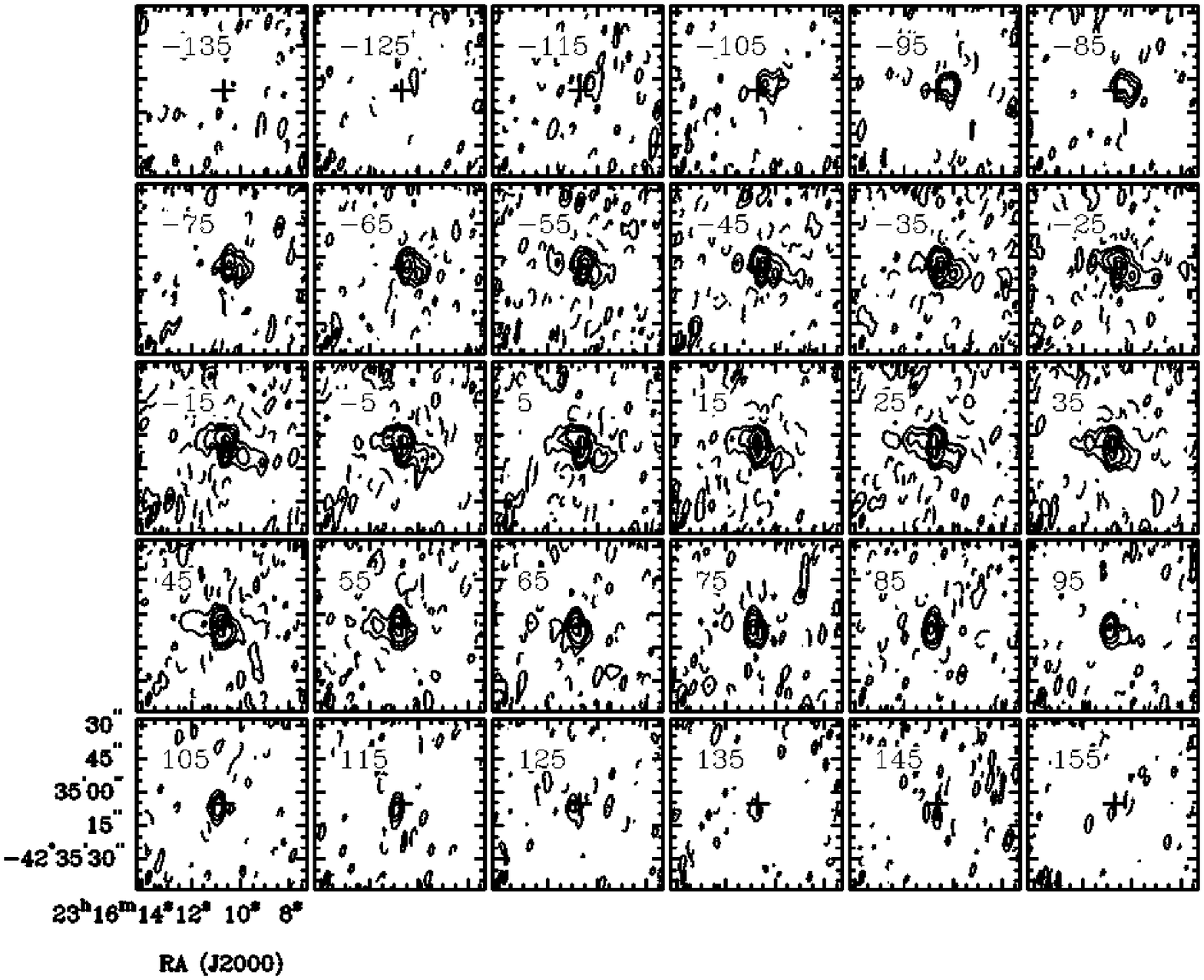}
\caption{Channel maps of the central region of NGC~7552 in $^{12}$CO~(2 -- 1). Contour levels are plotted at -2, 2, 5, 10, 25, 40, and $55 \times 78 {\rm \  mJy \ beam^{-1}}$ (rms noise level).  The velocity of each channel, in units of km s$^{-1}$, is shown in the upper left corner of each panel.    A cross is plotted at the center of the galaxy in each panel.  The synthesized beam is $7\farcs0 \times 2\farcs8$ at a position angle of -11\fdg9, and is shown at the lower left corner of each panel.\label{FIG_12CO.CM}}
\end{figure}

Figure~\ref{FIG_12CO.CM} shows our channel maps of the $^{12}$CO~(2 -- 1) line.  By contrast with the simplicity of the $^{13}$CO~(2 -- 1) channel maps, three components are now discernible: a compact central component that is the counterpart of that seen in $^{13}$CO~(2 -- 1), and two extensions on opposite sides of the center with one arm extending east from north and the other arm west from south.  The compact central component 
is detectable over the velocity range from about $-125 {\rm \ km \ s^{-1}}$ to $135 {\rm \ km \ s^{-1}}$ with respect to the systemic velocity, somewhat larger than the detectable velocity range of its counterparts in $^{13}$CO~(2 -- 1), HCN~(1 -- 0), and HCO$^{+}$~(1 -- 0).  
By contrast, both the eastern and western arm-like components are detectable over a narrower velocity range than the compact central component in any of the observed molecular lines.  The western arm is detectable over the velocity range from about $-85 {\rm \ km \ s^{-1}}$ to $55 {\rm \ km \ s^{-1}}$ and is therefore preferentially blueshifted, whereas the eastern arm is detectable over the velocity range from about $-55 {\rm \ km \ s^{-1}}$ to $75 {\rm \ km \ s^{-1}}$ and is therefore preferentially redshifted.

\begin{figure}
\plottwo{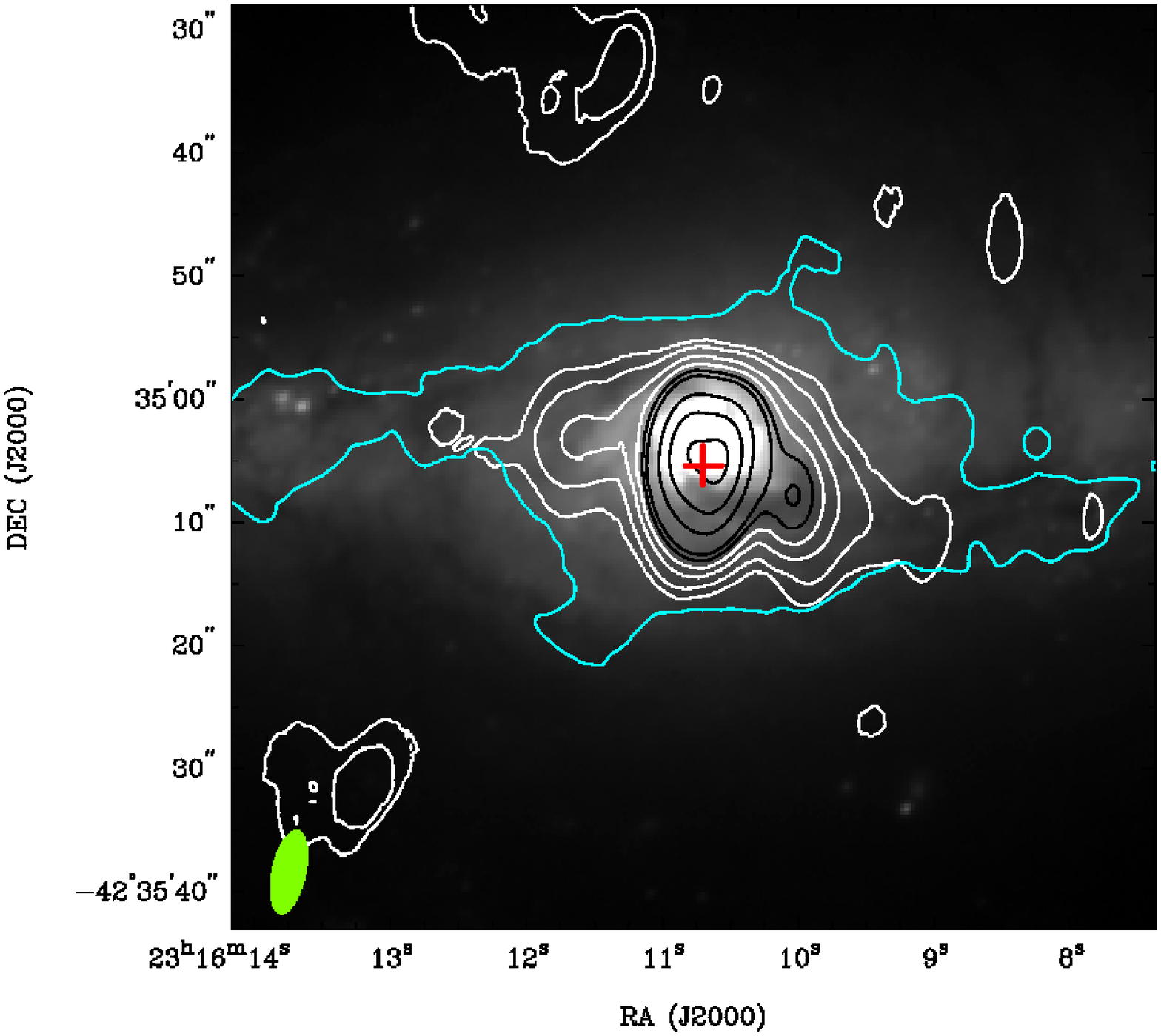}{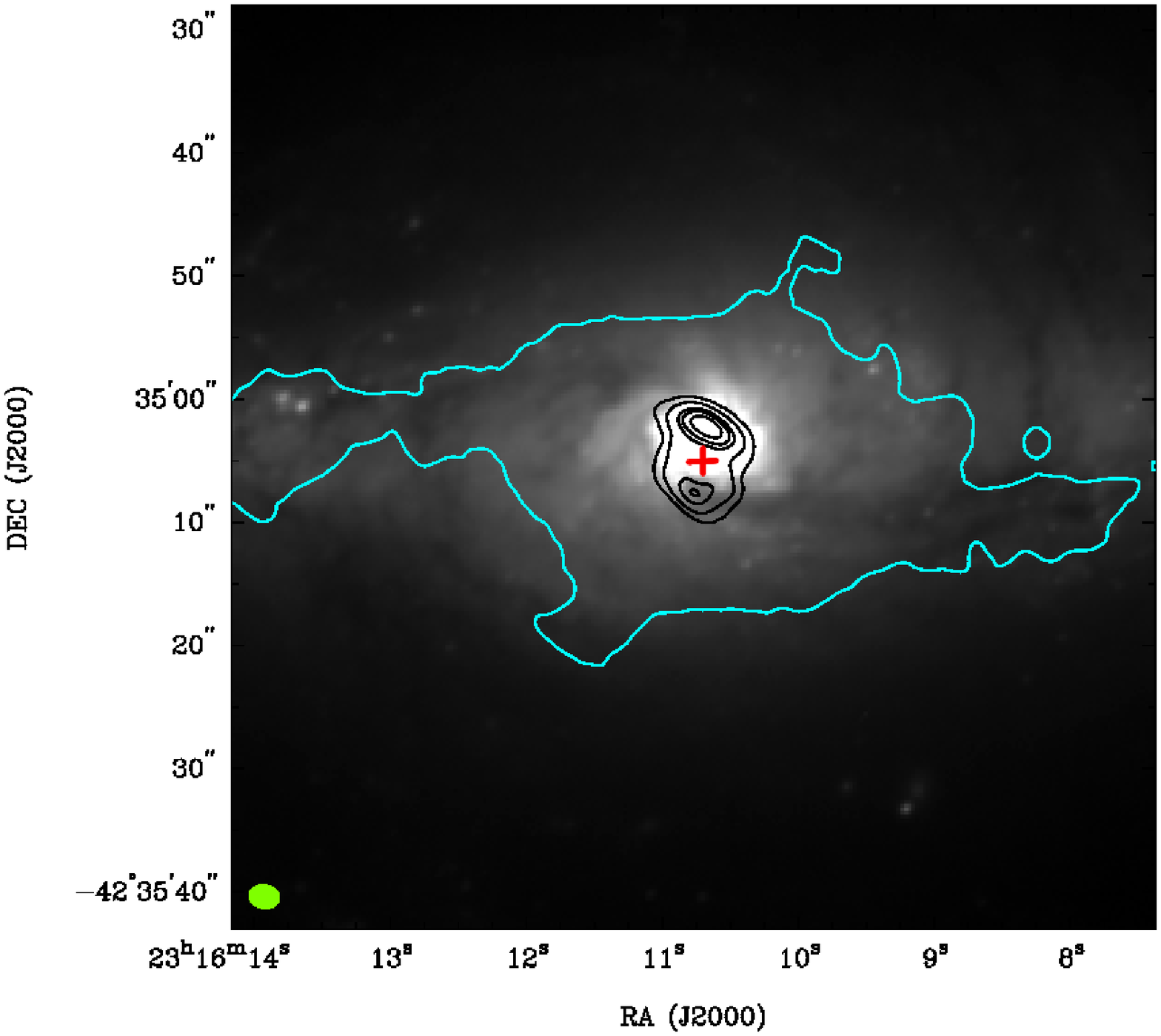}
\plottwo{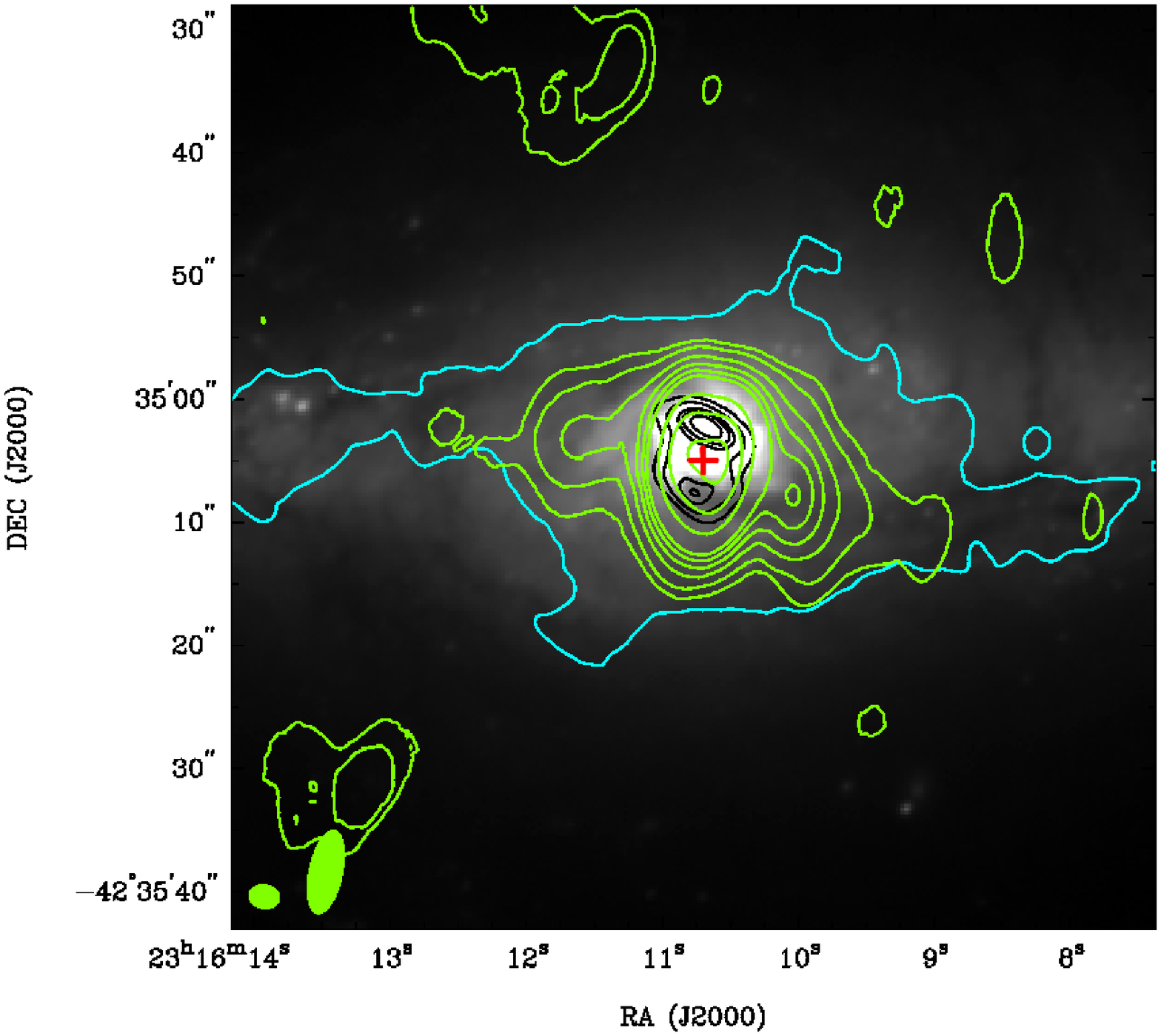}{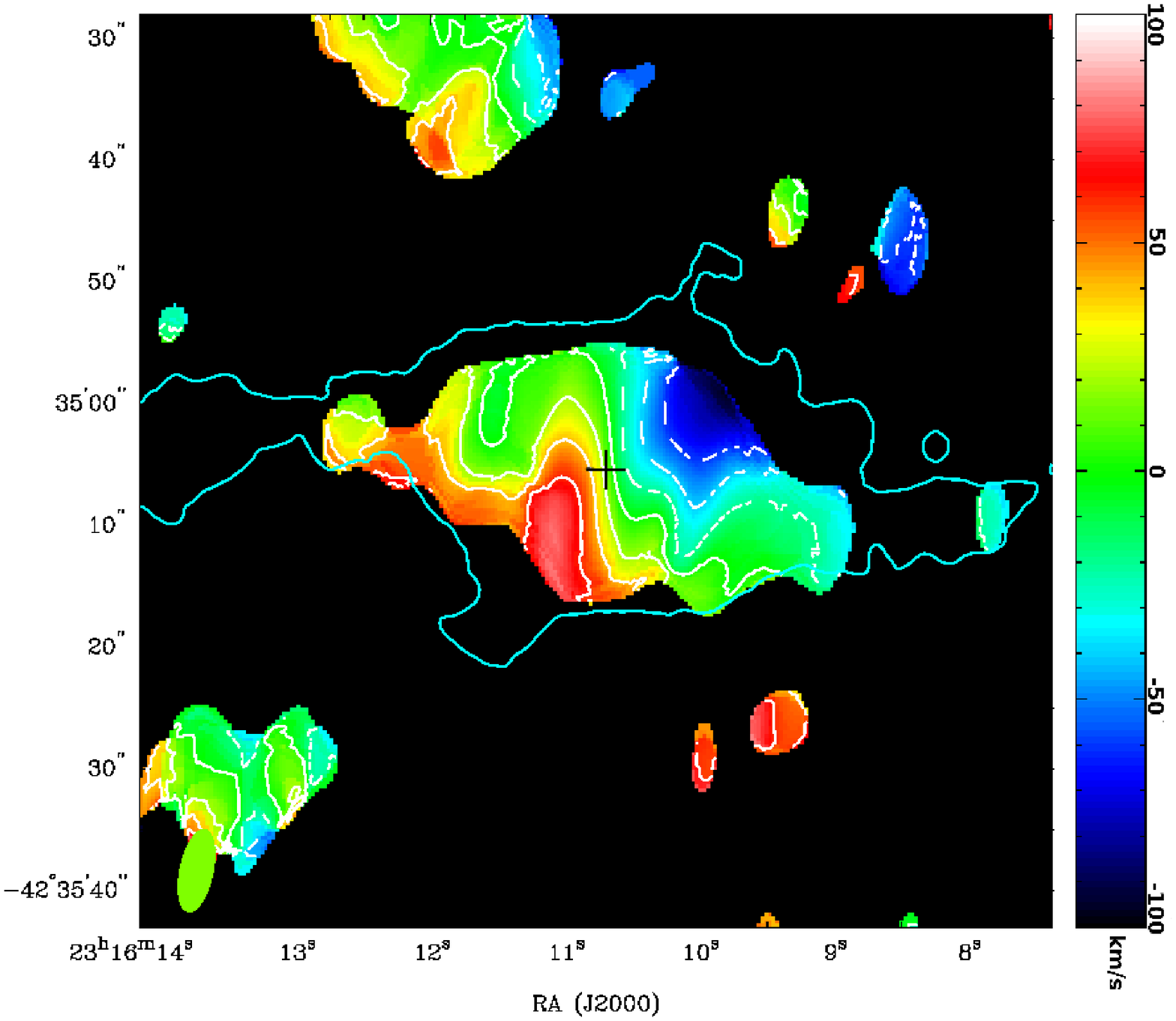}
\caption{Upper left: Integrated $^{12}$CO~(2 -- 1) intensity map in white/black contours plotted at 2, 5, 10, 15, 20, 25, 45, 65, and $85 \times 4.5 {\rm \ Jy \ beam^{-1} \ km \ s^{-1}}$ overlaid on the optical $i^{\prime}$-band image in grayscale shown in Figure~\ref{FIG_Optical}.  Upper right: Contours of the integrated HCN (1 -- 0) intensity map (contour levels are the same as those shown in Figure~\ref{FIG_HCN_MOM}) overlaid on the greyscale $i^{\prime}$-band image.  Lower left: Integrated $^{12}$CO~(2 -- 1) (green contours) and HCN (1 -- 0) (black contours) intensity maps both overlaid on the $i^{\prime}$-band image (grayscale).  Lower right: Intensity-weighted mean $^{12}$CO~(2 -- 1) velocity in color scale, as well as white contours plotted in steps of $20 {\rm \ km \ s^{-1}}$ starting from $-60 {\rm \ km \ s^{-1}}$ and ending at $60 {\rm \ km \ s^{-1}}$ with respect to the systemic velocity of 1586 km s$^{-1}$.   In each panel, cyan contour is the outline of the 5.8~$\mu$m emission at 5.5 MJy sr$^{-1}$ from the $Spitzer$ SINGS project published in \citet{Ken03}; this emission traces the dust lane visible in silhouette in the optical image.  A cross is plotted at the center of the galaxy, and the synthesized beam shown at the lower left corner, of each panel.\label{FIG_12CO_MOM}}
\end{figure}

Figure~\ref{FIG_12CO_MOM} (upper left panel) shows a map of the integrated $^{12}$CO~(2 -- 1) intensity in black/white contours overlaid on the optical $i^{\prime}$-band image of the inner part of the galaxy.  Dust lanes are visible as silhouettes in the optical image running east and west of the center, as well as in emission at 5.8~$\mu$m as shown by the cyan contour based on the observations by \citet{Ken03} in their {\em Spitzer} SINGS project.  The eastern and western $^{12}$CO~(2 -- 1) arms coincide with the innermost regions of these dust lanes.  These arms appear to connect with the circumnuclear starburst ring at or near the locations where we detected two knots of dense molecular gas as traced in HCN~(1 -- 0) and HCO$^{+}$~(1 -- 0) (see $\S$\ref{subsec_hcn_result}),  
as shown in the lower left panel of Figure~\ref{FIG_12CO_MOM} where the integrated $^{12}$CO~(2 -- 1) intensity is plotted in green contours and integrated HCN (1 -- 0) intensity in black contours.  We note that, for the same ratio in intensity between the arms and compact central component, the stronger western arm should be detectable (at 4.5$\sigma$) at the sensitivity of our HCN~(1--0)/HCO$^{+}$~(1--0) maps.

Figure~\ref{FIG_12CO_MOM} (lower right panel) shows a map of the intensity-weighted mean $^{12}$CO~(2 -- 1) velocity.  At our angular resolution, the kinematics of the compact central component cannot be completely separated from the individual kinematics of the eastern and western arms, and vice versa.  Nevertheless, the S-shape velocity field (contours) indicates the existence of non-circular motion, which is similar to what is observed in other barred galaxies.  In Figure \ref{FIG_PV_4lines}, we show a P-V diagram made from the $^{}$CO~(2 -- 1) channel maps at a position angle of $110\degr$.  As can be seen, the kinematics of the compact central component are similar to its counterparts in the other observed molecular lines.  The western arm is preferentially blueshifted whereas the eastern arm preferentially redshifted, and therefore both arms share the same overall kinematics as the large-scale galactic disk seen in \ion{H}{1}.

We derive the mass of molecular hydrogen gas in all the three features detected in $^{12}$CO~(2 -- 1) in the standard manner using the Galactic conversion factor between CO intensity and mass of molecular hydrogen gas of $X_{\mathrm{CO}}$ = 3 $\times$ 10$^{20}$ cm $^{-2}$ (K km s$^{-1}$)$^{-1}$ \citep{Sol87}.  This conversion is based on the $^{12}$CO~(1 -- 0) line, and in the present situation requires knowledge of the ratio in $^{12}$CO~(2 -- 1) to $^{12}$CO~(1 -- 0) intensities.  \citet{Aal95} find a ratio in brightness temperatures between these two lines of $1.2 \pm 0.1$ based on single-dish observations with the Swedish-ESO 15-m Submillimeter Telescope (SEST) towards the center of NGC~7552.  For simplicity, we shall therefore assume a ratio in brightness temperature of unity for $^{12}$CO~(2 -- 1) to $^{12}$CO~(1 -- 0) throughout this work.  Accordingly, the compact central component (corresponding to the circumnuclear starburst ring, and perhaps regions within), which has an integrated intensity in $^{12}$CO~(2 -- 1) of  $1487.1 \pm 20.3 {\rm \ Jy \ km \ s^{-1}}$, has a gas mass of $(3.2 \pm 0.1) {\times} 10^{9} {\rm \ M_{\odot}}$, including the mass of the heavy elements other than molecular hydrogen ($M_{\mathrm{gas}}=1.36M_{\mathrm{H_{2}}}$).  The integrated $^{12}$CO~(2 -- 1) intensity detected over the entire region is $2330.5 \pm 34.7 {\rm \ Jy \ km \ s^{-1}}$, and hence the mass of molecular gas in the eastern and western arms combined is $(1.7 \pm 0.2) {\times} 10^{9} {\rm \ M_{\sun}}$. 

The dense gas fraction can be estimated by comparing the mass derived from $^{12}$CO~(2 -- 1) and HCN (1 -- 0).  Assuming that the gas traced by HCN (1 -- 0) is gravitationally bound \citep{Sol92}, the HCN-to-H$_{2}$ conversion factor $X_{\mathrm{HCN}}\equiv N(\mathrm{H_{2}})/L'\mathrm{_{HCN}}$ is $\sim$20 $M_{\sun}$  (K km s$^{-1}$ pc$^{-2}$)$^{-1}$, where $N(\mathrm{H_{2}})$ is the column density of molecular hydrogen gas and $L'\mathrm{_{HCN}}$ the HCN~(1--0) luminosity.  For the measured $L'\mathrm{_{HCN}} \approx 7.5 \times 10^{7} \rm \ K \ km \ s^{-1} \ pc^{-2}$, the corresponding mass in dense gas, as traced in HCN~(1--0), is $\sim$$1.5 {\times} 10^{9} {\rm \ M_{\sun}}$.  The dense gas fraction is therefore about 50\% at the compact central component, much higher than is usually found in Galactic giant molecular clouds.  We note, however, that the inferred dense gas fraction may contain considerable uncertainty because the conversion factors $X_{\mathrm{HCN}}$ and $X_{\mathrm{CO}}$ used may not be appropriate for the circumnuclear regions of galaxies (although the use of different conversion factors can either raise or lower the dense gas fraction).  For example, based on virial arguments for the circumnuclear starburst ring of the nearby barred spiral galaxy IC~342, \citet{Mei01} argue that $X_{\mathrm{CO}}$ is only 1 $\times$ 10$^{20}$ cm $^{-2}$ (K km s$^{-1}$)$^{-1}$ or even smaller, in which case the dense gas fraction at the compact central component becomes nearly 100\%.

\section{Physical Properties of the Circumnuclear Molecular Gas}
\label{sec_physical_properties}
In this section, we derive the physical properties of the molecular gas in the circumnuclear starburst ring based on the ratio in brightness temperatures of the molecular lines observed.  We then investigate the dynamical stability of this gas.

\subsection{Emission Recovered in our Interferometric Observations}
\label{subsec_line_profile}
Line ratios inferred with an interferometer can be compromised if there exists uniformly-bright features with sizes beyond that detectable even by the shortest baseline. 
\citet{Aal95} have observed the central region of NGC 7552 with the SEST in $^{12}$CO~(1 -- 0), $^{12}$CO~(2 -- 1), $^{13}$CO~(1 -- 0), $^{13}$CO~(2 -- 1), and HCN~(1 -- 0). They do not show line profiles for any of the transitions observed, but only quote the line intensities or ratio in brightness temperatures of the different lines.  After convolving our channel maps to the same angular resolution as in the observation of the corresponding line with the SEST, we found that our SMA observation recovered $54\% \pm 9$\% of the $^{12}$CO~(2 -- 1) emission detected by the SEST. Given typical systematic uncertainties of about 20\% in the absolute flux calibration at millimeter wavelengths, we may well have recovered the bulk of the circumnuclear emission in $^{12}$CO~(2 -- 1).  Our ATCA observation recovered $76\% \pm 16\%$ of the HCN~(1 -- 0) emission detected by the SEST, suggesting once again that the bulk, if not all, of the circumnuclear emission in this line has been recovered.   
Surprisingly, our SMA observation recovered only $26\% \pm 10\%$ of the $^{13}$CO~(2 -- 1) emission detected by the SEST, despite the fact that $^{13}$CO~(2 -- 1) traces higher column densities and therefore more compact regions than $^{12}$CO~(2 -- 1) as evidenced by the maps presented in Figure~\ref{FIG_13CO_MOM}.  The $^{13}$CO~(2 -- 1) line however was detected at a confidence level of only 3--4$\sigma$ in the SEST observation, and we therefore do not place a high degree of confidence on this particular comparison.

\subsection{Physical Properties}
\label{sec_lineratio_LVG}
We computed the physical properties (density and temperature) of the molecular gas associated with the circumnuclear starburst ring from the ratio in brightness temperatures of the HCN~(1 -- 0) to $^{13}$CO~(2 -- 1) lines using the Large Velocity Gradient (LVG) approximation \citep[e.g.,][]{Gol74,Sco74}.  We used the $^{13}$CO~(2 -- 1) line because it is more likely than the $^{12}$CO~(2 -- 1) line to trace the same region as the HCN~(1 -- 0) line, as the maps we presented earlier indicate (see \S \ref{subsec_CO_result} and Figure \ref{FIG_13CO_MOM}).  HCN~(1 -- 0) is a tracer of high density molecular hydrogen gas;
as the $^{13}$CO~(2 -- 1) emission is much more optically thin than the $^{12}$CO~(2 -- 1) emission, the former is more strongly weighted towards regions of high column density and therefore also density.  Indeed, as we shall show, the $^{13}$CO~(2 -- 1) emission is optically thin and therefore a good tracer of column density.

To derive the ratio between the HCN~(1 -- 0) and $^{13}$CO~(2 -- 1) intensities, we first convolved our HCN~(1 -- 0) map to the same angular resolution as our $^{13}$CO~(2 -- 1) map (see Figure \ref{FIG_13CO_MOM}). The circumnuclear starburst ring is spatially unresolved at this angular resolution, and so the line ratio we derive for HCN~(1 -- 0) to $^{13}$CO~(2 -- 1) of $2.0 \pm 0.2$ is an average over the entire ring.  We assumed Galactic abundances for the $^{13}$CO and HCN molecules of  [$^{13}$CO]/[H$_{2}$]~$\approx$~$10^{-6}$ \citep[][]{Sol79} and [HCN]/[H$_{2}$]~$\approx$~$2 \times 10^{-8}$ \citep[][]{Irv87}.  Finally, we estimated a velocity gradient of $0.24 {\rm \ km \ s^{-1} \:pc^{-1}}$ from the P-V diagram along the kinematic major axis of the HCN~(1 -- 0) emission as shown earlier in Figure~\ref{FIG_PV_4lines}.

\begin{figure}
\epsscale{.90}
\plotone{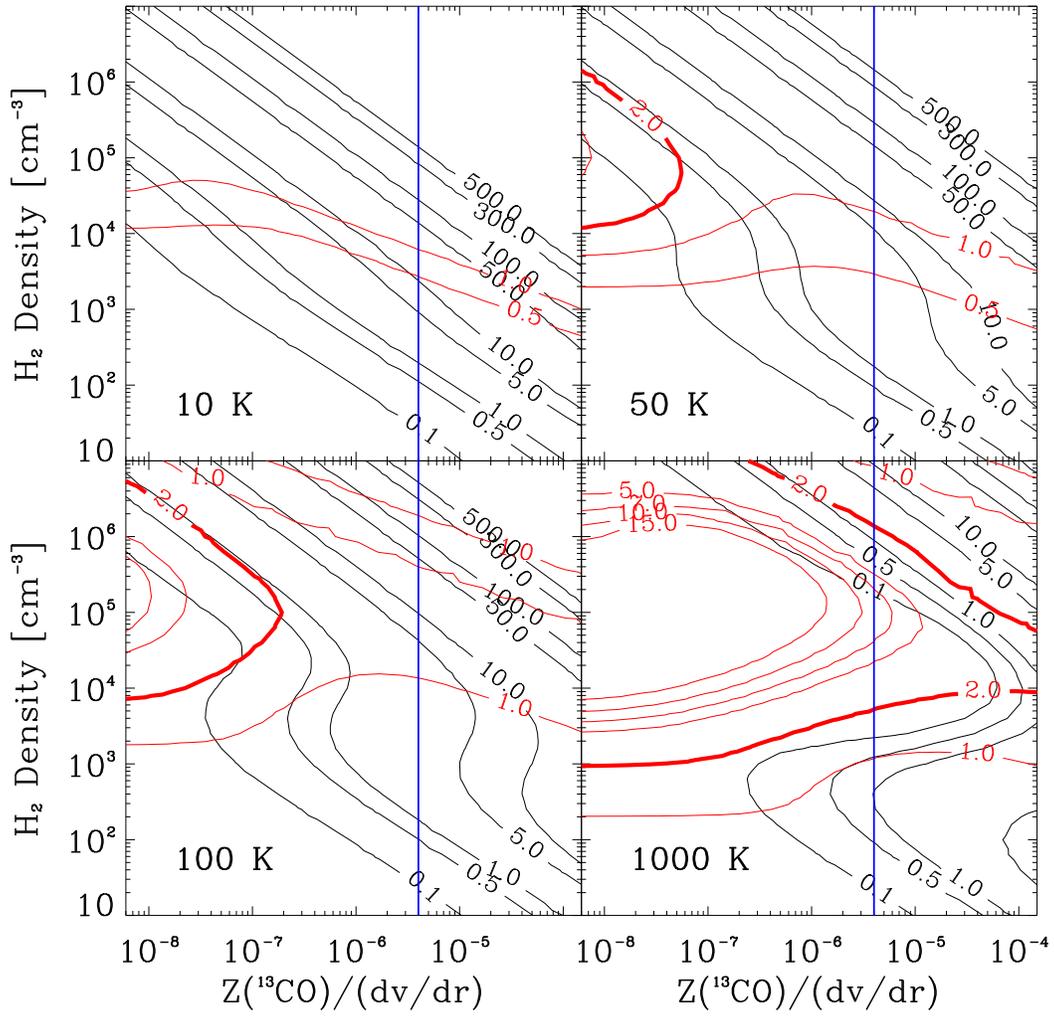}
\caption{Derived physical properties of the molecular hydrogen gas in the circumnuclear starburst ring based on our LVG analysis.  In all panels, density of molecular hydrogen gas is plotted as a function of the product $Z(^{13}$CO)/($dv/dr$), which for the present circumstances has an adopted value of $\sim$$4 \times 10^{-6}$ (vertical line in each panel; see text).  Black contours in each panel indicate the $^{13}$CO~(2 -- 1) opacity.   Red contours are the ratio in brightness temperature of HCN~(1 -- 0)  to $^{13}$CO~(2 -- 1), which averaged over the ring has a value of $2.0 \pm 0.2$ (thick red contour in each panel where visible).   The upper left panel are solutions for a kinetic temperature of 10~K; upper right panel 50~K; lower left panel 100~K; and lower right panel 1000~K.  As can be seen, for the measured line ratio and adopted product $Z(^{13}$CO)/($dv/dr$), solutions can only be found for kinetic temperatures well above 100~K.\label{FIG_LVG}}
\end{figure}

Figure \ref{FIG_LVG} shows a plot of the density of molecular hydrogen gas, $n(\rm H_2)$, as a function of the product $Z(^{13}$CO)/($dv/dr$), where $Z(^{13}$CO) is the $^{13}$CO abundance and $dv/dr$ the velocity gradient along the radial direction.  For the adopted or inferred values above, $Z(^{13}$CO)/($dv/dr$)~$\approx 4 \times 10^{-6}$, which is displayed with the blue vertical line in each panel.  The different panels in this figure correspond to different kinetic temperatures (from 10~K to 1000~K). The red contours in each panel indicate acceptable solutions in the $n(\rm H_2)$ and $Z(^{13}$CO)/($dv/dr$) phase space for different HCN~(1 -- 0) to $^{13}$CO~(2 -- 1) line ratios based on our LVG calculations.  The black lines indicate the $^{13}$CO~(2 -- 1) opacity.  As can be seen, for a HCN~(1 -- 0) to $^{13}$CO~(2 -- 1) line ratio of $2.0 \pm 0.2$, and $Z(^{13}$CO)/($dv/dr$)~$\approx 4 \times 10^{-6}$, solutions can only be found for kinetic temperatures well above 100~K, for which the $^{13}$CO~(2 -- 1) emission is optically thin.  At such temperatures, the density of the molecular gas (averaged over the line-emitting regions) is around 10$^{5\pm1}$ cm$^{-3}$. A similar result has been seen in the circumnuclear starburst ring of the barred spiral galaxy NGC~1097 \citep{Hsi08}.

On the other hand, the adopted Galactic abundance for $^{13}$CO of [$^{13}$CO]/[H$_{2}$]~$\approx$~$10^{-6}$ may be too high for starburst regions \citep{Tan98}.  The $^{13}$CO abundance is controlled primarily by fractionation reactions and photodissociation.  Fractionation reactions enhance the $^{13}$CO abundance, but such reactions are only effective in regions where the temperature is less than $\sim$35~K \citep{Wat76} and hence not exposed to vigorous star formation.   Photodissociation, naturally, is even more effective in regions where the star-formation rate is higher.  If the combination of these factors reduces [$^{13}$CO]/[H$_{2}$] by, say, an order of magnitude to $\sim$10$^{-7}$, then acceptable solutions to the observed HCN~(1 -- 0) to $^{13}$CO~(2 -- 1) line ratios can be found for temperatures $\sim$50--100~K and densities $\sim$10$^{5}$ cm$^{-3}$.  Thus, unless the $^{13}$CO abundance is even lower than the range of values considered above, the molecular gas that we detected in the starburst ring is relatively warm and dense, as would be expected for gas associated with regions that have recently undergone vigorous star formation \citep[see also][]{Mat98,Mat10}.

In many galaxies exhibiting nuclear activity, it is often difficult tell whether this activity is produced by active galactic nuclei (AGN) or starburst. \citet{Koh01} suggest that such galaxies can be distinguished by plotting the line ratio of HCN~(1 -- 0) to HCO$^+$~(1 -- 0) against that of HCN~(1 -- 0) to $^{12}$CO~(1 -- 0) (Figure \ref{FIG_HCN_diagram}).  They show that galaxies dominated by or exhibiting only AGN activity lie in the upper right corner of such a diagram, having high values in both line ratios.  By comparison, galaxies exhibiting only starburst activity or a mix of starburst and AGN activity lie in the lower left corner, having low values in both line ratios.  In the case of NGC~7552, we can clearly see that the nuclear activity is dominated by a starburst (i.e., circumnuclear starburst ring), and our observation are able to confidently separate the molecular gas associated with the starburst from any associated with the nucleus in HCN~(1 -- 0) and HCO$^+$~(1 -- 0), and also to some degree of confidence in $^{12}$CO~(2 -- 1) and $^{13}$CO~(2 -- 1) (based on the similarity of the gas kinematics in all four lines).  Such is not the case for the galaxies used to make the abovementioned plot in \citet{Koh01}.  For NGC~7552, the HCN~(1 -- 0) to HCO$^+$~(1 -- 0) line ratio, averaged over the circumnuclear starburst ring, is $1.13 \pm 0.04$.  The HCN~(1 -- 0) to $^{12}$CO~(1 -- 0) ratio, assuming that the $^{12}$CO~(1 -- 0) is equal to the $^{12}$CO~(2 -- 1) intensity for the reason mentioned earlier (see \S\ref{subsec_line_profile}), is $0.181 \pm 0.005$. This places NGC~7552 firmly in the region occupied by galaxies with nuclear activity dominated by a starburst, enhancing the validity of the plot made by \citet{Koh01}.

\subsection{Dynamical Stability}
We examine whether the molecular gas in the circumnuclear starburst ring is dynamically stable according to the Toomre criterion \citep{Too64}.  For a disk to be stable against collapse due to its own gravity, the Toomre Q parameter, which is defined as the ratio of the critical surface gas density to the actual surface gas density, $\Sigma _\mathrm{crit}/\Sigma _\mathrm{gas}$, is required to have a value equal to or exceeding unity.

We compute the average surface gas density in a circular region of radius of 10\arcsec\ ($\sim$1~kpc), just encompassing the circumnuclear starburst ring, using the molecular gas mass derived in \S\ref{subsec_CO_result} to find $\Sigma _\mathrm{gas} \approx 900 {\rm \ M_{\sun} \ pc^{-2}}$.  Because the $^{12}$CO~(2 -- 1) emission likely exhibits a central depression or hole within the circumnuclear starburst ring, the value we computed is likely to be a lower limit.
We estimated the critical surface gas density using the relationship from \citet{Koh99}
\begin{equation}
\Sigma _{\rm crit}=630\left (\frac{\sigma }{15 {\rm \ km \ s^{-1}}} \right )\left (\frac{\kappa }{0.57 {\rm \ km \ s^{-1} pc^{-1}}} \right ) {\rm \ M_{\sun} \ pc^{-2}} 
\label{EQ_SIG_CRIT2}
\end{equation}
where ${\sigma}$ is the one-dimensional velocity dispersion, for which we estimated a value of $12 {\rm \ km \ s^{-1}}$ based on a second moment map of the $^{12}$CO~(2 -- 1) emission (not shown), and $\kappa$ the epicyclic frequency.  As shown below in $\S$\ref{subsec_dynamical_resonances}, we derived an epicyclic frequency of $0.4 {\rm \ km \ s^{-1} \: pc^{-1}}$ at the radius of the circumnuclear starburst, and therefore $\Sigma _{\rm crit} \approx 354 {\rm \ M_{\sun} \ pc^{-2}}$.  The Toomre Q parameter is therefore no larger than 0.4 in the circumnuclear starburst ring, indicating that the molecular gas there is (highly) unstable to gravitational collapse. 
\section{Global Kinematics}
\label{sec_rotation_curve_kinematics}
In this section, we derive the global kinematics (rotation curve) of NGC~7552 and, assuming a pattern speed for the galactic bar, estimate the predicted locations of dynamical resonances.  We then compare these predictions with the locations of prominent features in NGC~7552 that in other disk galaxies are commonly attributed (but seldom demonstrated explicitly to correspond) to dynamical resonances; in the case of NGC~7552, \citet{Fo94b} have highlighted features that possibly correspond to dynamical resonances.
Our overarching goal is to determine whether we can understand the global kinematics of NGC~7552 sufficiently well so as to test ideas for the formation of dense molecular gas and stars in its circumnuclear starburst ring as discussed in the next section.
\subsection{Rotation Curve}
\label{subsec_brandt_rotation_curve}

As pointed out in $\S$\ref{subsec_HI_result}, the intensity-weighted mean \ion{H}{1} velocity map (Fig.~\ref{FIG_HI_MOM}) exhibits an S-shaped asymmetry characteristic of periodic orbits predicted theoretically for barred galaxies \citep[e.g.,][and references therein]{Ath92}. In general, such orbits are elliptical (and centered on the dynamical center of the galaxy), and have their major axes either parallel ($x_1$ orbits) or orthogonal ($x_2$ orbits) to the major axis of the bar, changing from one to the other between the locations of dynamical resonances.  Specifically, beyond the ILR (inner Lindblad resonance) out to corotation (CR), orbits are $x_1$.  Within the ILR, orbits are $x_2$.  If there is both an outer ILR (oILR) as well as an inner ILR (iILR), orbits are $x_1$ inside the iILR and $x_2$ between the iILR and oILR.  Between CR and the outer Lindblad resonance (OLR), orbits are $x_1$.  Close to and beyond the major axis of the bar (predicted theoretically to terminate at or before CR), both $x_1$ and $x_2$ orbits become progressively more (and are predicted to be usually close to) circular. Close to a bar potential, $x_1$ is the predominant orbit, hence the major kinematic axis is closely aligned with the galactic bar.  The change from $x_1$ to $x_2$ orbits and back between dynamical resonances causes the S-shaped twist in the major kinematic axis on scales up to that comparable with the length of the galactic bar. 

To accurately derive the rotation curve of NGC~7552 from our measurements of the projected galaxy kinematics, we require {\it a priori} knowledge of the axial ratios (which change with galactic radius) of the $x_1$ and $x_2$ orbits, and the inclination of the galaxy (including any variation in inclination with galactic radius if the disk is warped).  As these parameters (apart from perhaps inclination) are not in general known, 
we assumed a constant inclination of 25\degr\ for NGC~7552 based on the axial ratio of its stellar disk as described in $\S$1.  For the major kinematic axis of the disk, we used a position angle of 110\degr\ as indicated by the first moment map in all four molecular lines as well as the inner region of the first moment map in the \ion{H}{1} line.  

The HCN~(1 -- 0) to HCO$^+$~(1 -- 0) emissions originate primarily, if not entirely, from the circumnuclear starburst ring, and are therefore useful for inferring the rotational velocity of this ring.  Sampling only a limited and furthermore spatially unresolved radial range, however, these lines do not provide a meaningful rotation curve at the center of the galaxy.  The $^{12}$CO~(2 -- 1) line samples a larger radial range from (presumably) the ring to the innermost regions of the dust lanes, and is useful for making a crude estimate of the rotation curve over this range.  The \ion{H}{1} line provides a relatively robust estimate of the rotation curve beyond the outer regions probed in $^{12}$CO~(2 -- 1) to far beyond the stellar optical disk. 

The rotation curve is directly derived from P-V diagrams of $^{12}$CO~(2 -- 1) and \ion{H}{1} from cuts along 110\degr, as shown earlier in Figure \ref{FIG_PV_4lines}, with the envelope-tracing method \citep[see details and more examples in][]{Sof96,Sof97}. 
The method traces terminal velocity (or tangent-points) on PV diagram and corrects for the ISM velocity dispersion.
The deprojected rotational velocity is derived from the terminal velocity $V_{\mathrm{t}}$:  
\begin{equation}
V_{\mathrm{rot}}=V_{\mathrm{t}} /\sin i\: -\: \left ( \sigma _{\mathrm{obs}}^{2}\: +\:\sigma _{\mathrm{ISM}}^{2}  \right ),\label{EQ_Vter}
\end{equation}
where $i$ is the inclination, $\sigma _{\mathrm{obs}}$ the velocity resolution of the observation, and  $\sigma _{\mathrm{ISM}}$ the velocity dispersion of  interstellar gas.  We adopt an ISM velocity of 7 km s$^{-1}$ \citep{Sta89,Mal94}. The terminal velocity is the velocity at which the intensity ($I_{\mathrm{t}}$) is equal to:
\begin{equation}
I_{\mathrm{t}}=\left [ \left ( \eta I_{\mathrm{max}} \right )^{2}\: +\: I_{\mathrm{lc}}^{2} \right ]^{1/2}
\label{EQ_It}
\end{equation}
on the P-V diagram, where $I_{\mathrm{max}}$ and $I_{\mathrm{lc}}$ are, respectively, the maximum intensity and intensity of the lowest (typically the 3$\sigma$) contour level. 
If the intensity is high enough, $I_{\mathrm{t}}$ $\sim$ $I_{\mathrm{lc}}$ and is therefore approximately defined by the locus of the rotation curve.
$\eta$ is taken as 0.2 as suggested in \citet{Sof96}, corresponding to the 20\% level of the intensity profile at a fixed position, $i.e.$, $I_{\mathrm{t}}\simeq 0.2\times I_{\mathrm{max}}$. 
The accuracy of the obtained $V_{\mathrm{t}}$ along with the $V_{\mathrm{rot}}$ by the envelope-tracing method is $\sim$ $\pm$15 km s$^{-1}$, corresponding to $\sim$ $\pm$15 km s$^{-1}$/$\sin i$ in deprojected velocity \citep{Sof96,Sof97}. 

In Figure \ref{FIG_PV_RC}, we plot the projected rotation curve derived from the $^{12}$CO~(2 -- 1) (from 0 to $\sim$ 0.8 kpc)  and \ion{H}{1} (beyond 1 kpc) lines onto the P-V diagram of both these lines cut along a position angle of $110\degr$. 
The projected uncertainties of $\pm$15 km s$^{-1}$ are plotted on the approaching side of the rotation curve to indicate the size of the error bars.
Due to the imperfect physical connection between the two tracers and asymmetry of P-V diagrams, the rotation curve between 0.5 to 1 kpc is poorly connected; we indicate these regions with grey shading in Figure \ref{FIG_PV_RC}. 

The deprojected inner and the entire derived rotation curves are shown respectively in Figures \ref{FIG_RC1} and \ref{FIG_RC2}. The open and solid black circles denote the receding and approaching sides of the rotation curve respectively, and the red open circles indicate the average of the two sides at the radii where both receding and approaching velocities are seen. The deprojected uncertainties are plotted in the lower right of each panel.

\subsection{Dynamical Resonances}
\label{subsec_dynamical_resonances}
Because the intensity of the approaching side in the \ion{H}{1} map is significantly weaker than that of the receding side, together with the asymmetry of the derived rotation curve on each side, we determine the dynamical resonances for each side of the galaxy separately. 
We compute where dynamical resonances are theoretically predicted to occur using the usual assumption that the angular velocity of the bar (its pattern speed, $\Omega _{\mathrm{p}}$) is equal to the angular velocity of the galaxy at the outermost ends of the bar (i.e., the CR).  We determined the length of the bar in NGC~7552 from the Two Micron All-Sky Survey (2MASS) image at 2.2 $\mu$m, assuming that the bar ends where the two prominent spiral arms first appear to emerge at a radius of about 5~kpc. 

The average rotation curve shows that the angular velocity of the galaxy at the bar's end (5 kpc) is about $38 {\rm \  km \ s^{-1} \ kpc^{-1}}$, which we therefore take to be the pattern speed of the bar.  
In Figure~\ref{FIG_Resonance}, we plot the angular velocity (velocity divided by radius, $V$/$R$) of the galaxy, ${\Omega}$ (filled circles), as derived from the rotation curve.   The solid line is an interpolation between adjacent angular velocity data points, from which we derived the epicyclic frequency, $\kappa$, of the orbits, where $\kappa ^{2}=2\left ( \frac{v^{2}}{R^{2}}\; +\; \frac{v}{R}\frac{\mathrm{d}v }{\mathrm{d} R}\right )$.  We then computed the $\Omega - \kappa/2$ curve (triangles) to infer the location(s) of the ILRs, the $\Omega - \kappa/4$ curve (diamonds) to infer the location of the Ultra Harmonic Resonance (UHR), and the $\Omega + \kappa/2$ curve (squares) to infer the location of the OLR. The uncertainties in these curves are rather large so we do not show the error bars in Figure~\ref{FIG_Resonance} for clarity, but show the possible radial ranges of the resonances taking these uncertainties into account. 

We therefore constrain the following possible locations of dynamical resonances in NGC 7552. The $\Omega + \kappa/2$ curve (open squares) cross the value of $\Omega _{\mathrm{p}}$ at $>$8~kpc, the predicted location of the OLR.
Theory predicts that the OLR in spiral galaxies should occur just beyond the outermost reaches of their spiral arms, which in the case of NGC~7552 can be traced out to a (deprojected) radius of about 10~kpc.  
The computed $\Omega - \kappa/4$ curve implies that the UHR occurs at roughly 2 -- 4~kpc. \citet{Fo94b} suggest that an isolated bright \ion{H}{2} region located at a (deprojected) radius of 3.5~kpc coincides with the UHR. The $\Omega - \kappa/2$ curve implies two ILRs. The oILR is located at a radius of about 1.7~kpc, which is beyond the circumnuclear starburst ring that has a radius of about 0.5~kpc. Further inwards the angular velocity might eventually drop to a value of 0 km s$^{-1}$ (in principle) at the center of the galaxy, and so an iILR could exist somewhere within the circumnuclear starburst ring.  On the other hand, we do not exclude the possibility that the rotation curve keeps rising due to a black hole in the galactic center, resulting in the absence of an iILR.  Our finding that the circumnuclear starburst ring is located within the oILR and beyond, possibly, the iILR is consistent with theoretical predictions \citep{But96}.  The reasonable agreement between the computed and anticipated locations of dynamical resonances in NGC 7552 provides a measure of confidence in our understanding of the global kinematics of this galaxy.

\section{Star Formation in the Circumnuclear Ring}
\label{sec_sf}
In this section, we investigate the star formation properties of the circumnuclear starburst ring, determine whether the observed properties of the ring are consistent with the model proposed by \citet{Koh99}, and if not what modifications are required.

\label{sec_discussion_Star Formation Rate and Timescale}
\subsection{Star Formation Rate}
\label{sec_discussion_SFR}
The radio continuum observations of \citet{Fo94a}, the data from which we reprocessed here, provides the cleanest (i.e., not subject to extinction, emission detected at high S/N) measure of the star formation rate (SFR) in the circumnuclear ring.  Like \citet{Fo94a}, we used the standard prescription suggested by \citet{Con92} to derive the supernova rate as $\sim$0.1 yr$^{-1}$ from the measured radio luminosity for the ring.   We then estimated the SFR from the supernova rate in the manner also suggested by \citet{Con92}, but corrected for a Salpeter Initial Mass Function (IMF) as suggested by \citet{Con02}, as about $15 {\rm \ M_\sun \ yr^{-1}}$.  An independent estimate of the SFR in the ring can be made from hydrogen recombination lines in the optical and near-infrared, although such estimates are affected by extinction.   The luminosity of the H$\alpha$ line of 1.5 $\times$ 10$^{42}$ erg s$^{-1}$ derived from the Gemini image, using the relationship of $L_{\mathrm{H\alpha }}$ - SFR suggested by \citet{Ken98}, implies a SFR of about $12 {\rm \ M_\sun \ yr^{-1}}$.  \citet{Moo88} find a luminosity in the Br$\gamma$ line of 9.8 $\times$ 10$^{39}$ erg s$^{-1}$ for the circumnuclear ring.  Using the relationship between the Br-$\gamma$ line luminosity and SFR suggested by \citet{Ken98}, the SFR is about $8 {\rm \ M_\sun \ yr^{-1}}$.  The correlation between the global HCN~(1 -- 0) and far-infrared luminosities found by \citet{Gao04} for normal to extreme star-forming galaxies can be used to provide an estimate of the SFR using the measured HCN~(1 -- 0) luminosity ($L{}'_{\mathrm{HCN}}$) as a proxy for the infrared luminosity ($L_{\mathrm{IR}}$).  With an $L{}'_{\mathrm{HCN}}\approx7.5 \times 10^{7}$ K km s$^{-1}$ pc$^{2}$ and hence $L_{\mathrm{IR}}$ of  $(5.5 \pm 0.1) \times 10^{10} {\rm \ L_{\sun}}$ in the circumnuclear starburst ring, the corresponding star formation rate (for a Salpeter IMF) is 11.0 ${\pm}$ 0.2 M$_{\sun}$ yr$^{-1}$. Therefore, in summary, the SFR in the circumnuclear starburst ring of NGC 7552 is about 10--15 M$_{\sun}$ yr$^{-1}$.

For comparison, the global SFR derived in the same manner from observations of the entire galaxy in the radio continuum at 20~cm by \citet{Sch06} is about $18 {\rm \ M_\sun \ yr^{-1}}$.  
The global star formation rate derived from the H$\alpha$ line, based on our Gemini observations, is about $47 {\rm \ M_\sun \ yr^{-1}}$.   
The global star formation rate can also be estimated from the far-IR measurements of NGC~7552 with IRAS by \citet{San03}.  They find an infrared luminosity for this galaxy of $1.1 \times 10^{11} {\rm \ L_{\odot}}$, thus placing NGC~7552 right at the threshold for a Luminous Infrared Galaxy.  Using as above the conversion between infrared luminosity and SFR suggested by \citet{Gao04} for a Salpeter IMF, the SFR is about $22 {\rm \ M_\sun \ yr^{-1}}$. 
Because the SFR estimated from H$\alpha$ can be highly uncertain \citep[e.g., up to 99\% of UV photons can leak out of \ion{H}{2} regions;][]{Ken98}, we give more weight to the other methods of estimating SFR as  described above.
Thus, the circumnuclear region of NGC~7552 contributes $\sim$50 -- 80\% of the galaxy's entire star formation, and being strongly concentrated in the circumnuclear ring, results in an extremely high star formation rate surface density of $\gtrsim$$4 \times 10^{-6} {\rm \ M_\sun \ yr^{-1} \ \ pc^{-1}}$.

The SFR in the circumnuclear ring, divided by the mass of molecular gas present, is about (3 -- 5) $\times$ 10$^{-9}$ yr$^{-1}$. The inverse of this ratio represents the timescale for gas exhaustion and is $\sim$ 3 $\times$ 10$^{8}$ yr. Such a short gas exhaustion timescale is comparable to that of interacting objects and ULIRGs (few $\times$ 10$^{8}$ yr) and about 10 times shorter than that in normal galaxies \citep{Big08}. Note that even though the gas exhaustion timescale of circumnuclear starburst ring is short, as the galaxies with circumnuclear rings are generally harboring stronger bars than the galaxies lacking rings, the gas is likely to be nearly continuously replenished at the circumnuclear region through, presumably, the bar-driven gas inflow.

\subsection{Formation of Dense Molecular Gas and Stars at the Circumnuclear Ring}
\label{sec_discussion_Formation}

The most prominent optical features --- circumnuclear starburst ring, large-scale bar, and spiral arms --- in NGC~7552 are remarkably similar to those of NGC~6951, including even the orientation of these features on the sky, thus providing a simple and direct comparison.  Like NGC~7552, the bar in NGC~6951 is oriented east-west, with spiral arms emerging from the bar and curling north on the western side and south on the eastern side.  Dust lanes that run along the bar intersect the circumnuclear starburst ring at its northern and southern sides, just like those seen in NGC~7552.  

\citet{Koh99} reported observations of the circumnuclear starburst ring of NGC~6951 in both $^{12}$CO~(1 -- 0) and HCN~(1 -- 0) with resolution of $\sim$500 pc in both lines.  They found twin CO knots at or close to the location where the dust lanes meet the ring, and twin HCN knots displaced downstream (assuming that the spiral arms are trailing) from the CO knots.  On this basis, \citet{Koh99} suggested that the molecular gas at the locations where the dust lanes meet the ring --- the twin CO knots, comprising gas channeled inwards by the bar --- is too turbulent to form dense molecular clouds.  Instead, it is only when this gas moves from $x_1$ orbits (closely traced by the dust lanes) beyond the oILR to $x_2$ orbits within the oILR (the circumnuclear starburst ring)  that dense molecular gas --- seen as the twin HCN knots --- are able to form; the velocity dispersion along $x_2$ orbits is predicted to be much smaller than that along $x_1$ orbits, and hence the molecular gas along $x_2$ orbits is much more susceptible to gravitational instabilities (see their Fig.~17). This scenario is not supported in our observation at least in a global sense as we find that the molecular gas in the circumnuclear starburst ring is very unstable to collapse. 

\citet{Kri07} have observed the circumnuclear starburst ring of NGC~6951 at significantly higher angular resolution ($\sim$240 pc) and sensitivity also in $^{12}$CO~(2 -- 1) and HCN~(1 -- 0).  An examination of their results (see their Fig.~1) reveals no discernible displacement between the centroids of the CO knot and dominant HCN knot in the northern part of the ring, and at best a small displacement between the centroids of the CO knot and dominant HCN knot in the southern part of the ring.  (There is a weaker HCN knot located just downstream of each dominant HCN knot: together, these knots, when convolved to the lower angular resolution attained in the observation by \citet{Koh99}, produce an apparent angular displacement between the HCN and CO knots.)  The centroids of all these knots are located at or close to where the dust lanes meet the ring, just as we see for the two HCN knots in the circumnuclear starburst ring of NGC~7552.  It is possible that collisions between gas clouds at the intersection of $x_1$ orbits (closely traced by the dust lanes) and the $x_2$ orbits (the circumnuclear starburst ring) give rise to relatively dense molecular gas \citep[e.g.,][]{Com85}, although the relative collision velocity must not be so high ($\lesssim 50 {\rm \ km \ s^{-1}}$) as to dissociate or ionize the molecular gas.  Alternatively, or in addition, turbulence generated by shocks help compress the molecular gas \citep{Elm94}; turbulence can help support giant molecular clouds against collapse on large scales, and promote the formation of dense condensations --- from which stars form --- on small  scales.

As sketched in Figure~\ref{FIG_revised_model}, both our observations of NGC~7552 and that by \citet{Kri07} of NGC~6951 therefore suggest an important revision to the picture proposed by \citet{Koh99}.  In both these galaxies, we suggest that collisions between diffuse (CO) molecular gas being channeled inwards along the dust lanes, turbulence generated by shocks, along with gravitational instabilities perhaps help promote the formation of dense molecular gas at or close to the intersection of the $x_1$ and $x_2$ orbits where the dust lanes meet the circumnuclear starburst ring.  Indeed,  for NGC 7552, the ratio in brightness temperature of the HCN~(1 -- 0) to $^{12}$CO~(2 -- 1) emission averaged over the ring is $\sim$0.18, whereas that at the stronger $^{12}$CO~(2 -- 1) western arm has a 3$\sigma$ upper limit of $\sim$0.12.  This indicates that the formation of dense molecular gas is promoted where the dust lanes meet the ring, relative to regions immediately beyond, along the same (at the very least for the western) dust lanes.

Our revised picture also helps explain another observed property of the circumnuclear starburst ring in both NGC~7552 an NGC~6591, and that is the displacement between the centroids of their HCN and radio continuum knots.  Such a displacement was pointed out above for NGC~7552 (see $\S$\ref{subsec_hcn_result} and Figure \ref{FIG_HCN_MOM}), and is obvious even through a cursory visual comparison of the molecular line maps made by \citet{Kri07} with the radio continuum maps made by \citet{Sai02} in NGC 6951. 
If we assume that the spiral arms in NGC~7552 are trailing as assumed by \citet{Koh99} for NGC~6951, then the radio continuum knots are located downstream of the HCN knots in both NGC~7552 and NGC~6951.  The most straightforward explanation for this displacement is that it reflects the timescale between the formation of massive stars from dense molecular gas at or close to the intersection of the $x_1$ and $x_2$ orbits, and the eventual demise of these stars in supernova explosions.  This argument can be checked by calculating the time it would take for the rotation of the circumnuclear starburst ring to have carried the radio continuum knots downstream away from the present location where the dust lanes meet the ring (specifically, the present location of the HCN knots), keeping in mind that during this time the pattern speed of the bar also carries the dust lane downstream with respect to the ring, albeit by a smaller angular distance.  

Figure \ref{FIG_Polar_Coor} shows the 3-cm radio continuum map of Figure~\ref{FIG_Continuum} and HCN integrated intensity map of Figure \ref{FIG_HCN_MOM} plotted in polar coordinates.  The northern radio and HCN knots have a deprojected angular offset $0\farcs47 \pm 0\farcs05$ ($\sim$50 pc), and the southern radio and HCN knots a deprojected angular offset of $0\farcs67 \pm 0\farcs08$ ($\sim$70 pc). 
As shown in $\S$\ref{subsec_brandt_rotation_curve}, the average rotational velocity of both sides of the circumnuclear starburst ring is about $218 {\rm \ km \ s ^{-1}}$, and the pattern speed of the bar about $38 {\rm \ km \ s^{-1} \ kpc^{-1}}$, corresponding to $19 {\rm \ km \ s^{-1}}$ at the ring (0.5 kpc); their net velocity is therefore about $199 {\rm \ km \ s ^{-1}}$.  
The propagation time between the northern radio and HCN knots is therefore (2.4$\pm$ 0.3) $\times$ 10$^{5}$ year, and that between the southern radio and HCN knots (3.4 $\pm$ 0.4) $\times$ 10$^{5}$ year.  
These timescales are about an order of magnitude shorter than the 5--10~Myr timescale for OB stars to explode as supernovae.  It is possible that the true pattern speed is significantly higher and/or the rotational velocity of the ring significantly slower than what we inferred, although it seems unlikely that both these speeds have been underestimated by up to an order of magnitude.  

In the case of NGC~6951, the centroids of the twin radio continuum knots imaged by \citet{Sai02} coincide closely with the centroids of the twin HCN knots as imaged by \citet{Koh99}, who estimated a propagation time between their observed twin CO and HCN knots of a few million years.   This is in better agreement with, although as in NGC~7552 on the shorter side of, the timescale for the demise of massive stars with a Salpeter-like IMF.  We point out that the model proposed above for the formation of dense gas and stars in the circumnuclear starburst ring can be more cleanly applied to NGC~6951 than NGC~7552.  The latter exhibits not just two but five radio continuum knots in its circumnuclear starburst ring (see Fig~\ref{FIG_Continuum}, right panel), indicating that the actual pattern for star formation --- and by extension formation of dense molecular gas --- is more complicated than can be fully captured by the simple model proposed in Figure~\ref{FIG_revised_model}.  In reality, dense molecular gas may form in the dust lanes, as they have been found in some bar galaxies with HCN (1 -- 0) observations \citep[e.g.,][]{Leo08}, before being channeled into the circumnuclear starburst ring (albeit, as shown in the case of NGC~7552, not as efficiently as at the intersection of these dust lanes with the ring), and continue to form in the circumnuclear starburst ring downstream from where the dust lanes meet the ring (e.g., through gravitational instabilities).

Even in the simpler case of NGC~6591, let alone NGC~7552, there are further complications that need to be taken into account. The formation of dense molecular gas at any given location may not proceed at a constant rate but is modulated in time. Some star formation may therefore occur even before gas is channeled into the ring and continue further downstream away from where the dust lanes meet the ring.   
In addition, we are not able to separate thermal and non-thermal contributions to the radio continuum at this stage, and thus cannot attribute all the emission to a post-supernova stage. \citet{Con92} suggests a simple way to separate the thermal emission from the total observed radio flux at a given frequency based on the spectral index.  For a spectral index of --0.89 derived in the \S\ref{subsec_continuum_result}, the fraction of thermal emission is about 40\% at 3~cm ($\sim$10 GHz).   If thermal (i.e., free-free emission from \ion{H}{2} regions) is enhanced at or just downstream of the HCN knots due to star formation, then the convolution (due to the limited angular resolution of our observation) of this emission with non-thermal emission produced further downstream by supernovae would tend to shift the centroid of the combined emission upstream towards the HCN knots.  The result is a smaller apparent propagation time between the HCN and radio continuum knots than is actually the case. On the other hand, the bright millimeter lines seen at the contact points may reflect high velocity dispersions \citep{Dow93} and therefore large linewidths, and the concentrations of dense molecular gas actually peak somewhat downstream of the HCN knots.  In this case, the result is a larger apparent propagation time between the HCN and radio continuum knots than is actually the case.
Finally, the IMF for star formation in the ring may be top-heavy, resulting in a shorter than usual timescale between the formation and death of the majority of massive stars (specifically, the time at which the radio intensity of all their supernovae combined peaks following the formation of these stars at the same epoch) compared with a Salpeter-like IMF.  \citet{Pap10} and \citet{Pap11} have pointed out that, in regions of high star-formation densities, the resulting high cosmic-ray densities can heat molecular gas to much higher temperatures than is the case in Galactic GMCs, in agreement with the relatively high temperatures than we infer for the molecular gas in the starburst ring of NGC~7552.  The higher temperature of the molecular gas requires a correspondingly higher characteristic Jeans mass for star formation in this gas, resulting in: (i) stars with larger masses than are found in Galactic GMCs; and (ii) a top-heavy IMF as low-mass star formation is suppressed \citep{Pap11}.

Our revised model of the sequence of star formation in this circumnuclear ring is consistent with the $pearls\;on\;a\;string$ scenario of \citet{Bor08} (see their Figure 7), which is based on their near infrared (H- and K-band) observations along with the age identification using the intensity ratios of emission lines. The $pearls\;on\;a\;string$ model proposes that the overdense region is close to where the gas enters the ring, and therefore where the starburst is triggered. The star clusters formed subsequently will continue their orbit along the circumnuclear ring. In this scenario, a bi-polar age gradient is expected, as has been seen in some galaxies, based on infrared and optical studies, such as NGC 1068 \citep{Dav98}, NGC 7771 \citep{Smi99}, M100 \citep{Ryd01,All06}, IC 4933 \citep{Ryd10}, and M83 \citep{Kna10}.

The infrared age dating method has been applied to NGC 7552 by \citet{Bra12}, who identify nine unresolved mid-infrared peaks (star clusters) in the starburst ring. There are tentative signs of an age gradient, in that the relatively young clusters are located near the north contact point, while the older clusters are found more than a quarter of the way around the ring away from the contact points. This result is qualitatively consistent with the age gradient expected by the $pearls\;on\;a\;string$ scenario and our revised model, although it must be noted that the age difference among the clusters is rather small at between only 5.5 and 6.3 Myr \citep[see Figure 10 in][]{Bra12}.

Even though the aforementioned bi-polar age gradients are found in the circumnuclear ring surveys of \citet{Bor08} and \citet{Maz08}, the detection rate of such age gradient patterns is only 60\% and 43\%, respectively, in these two surveys.   We cannot therefore rule out the possibility of other scenarios for star formation, in addition to that described above, in the circumnuclear starburst ring of NGC~7552. For example, \citet{Bor08} suggest a second $Popcorn$ model in which the gas clouds become gravitationally unstable due to turbulence \citep{Elm94}, and collapse at more random times and locations. Accordingly, there is no clear age gradient expected in the $Popcorn$ model, as found by \citet{Hsi11} for the circumnuclear starburst ring of NGC~1097.

In NGC~7552, we detect five radio continuum knots, not just two.  In NGC~6591, apart from the two dominant HCN(1 -- 0) peaks, there are two other HCN(1 -- 0)  knots in the circumnuclear starburst ring.  Thus, even for NGC~7552 and NGC~6591, the overall picture for the formation of dense molecular gas and stars in circumnuclear starburst rings cannot be as simple as that sketched in Figure~\ref{FIG_revised_model}. 
In reality, the formation of dense molecular gas may be promoted at the contact points, but also elsewhere in the circumnuclear starburst ring and star formation also occur there.  Both the $pearls\;on\;a\;string$ and $Popcorn$ models may operate simultaneously in any given circumnuclear starburst ring, complicating any single interpretation. What we present here is a zeroth-order model; the true picture requires observations at higher angular resolutions to properly study all locations where dense molecular gas is forming.

\label{sec_discussion_Formation}

\section{Summary and Conclusions}
\label{sec_summary}
NGC~7552 is a barred galaxy that, like a significant fraction \citep[$\sim$20\%--50\%;][]{Kna05} of such galaxies, exhibits a circumnuclear starburst ring.  A pair of dust lanes can be seen running along the bar on opposite sides of the center of the galaxy, and at their inner ends connect with the northern and southern parts of the circumnuclear starburst ring.  The circumnuclear ring currently contributes 50\% -- 80\% of the galaxy's entire star formation, and has a gas consumption timescale intermediate between normal and extreme star-forming (ultraluminous-infrared) galaxies.  

To better understand the sequence and circumstances under which dense molecular gas and stars form in this ring, we have observed the center of NGC~7552 in molecular lines tracing relatively diffuse molecular gas ($^{12}$CO and $^{13}$CO $J$ = 2 -- 1), which is the characteristic of Galactic giant molecular clouds; and relatively dense molecular gas (HCN and HCO$^+$ $J$ = 1 -- 0), which is the characteristic of dense condensations in giant molecular clouds where stars form.  We find that emission from:

\begin{itemize}

\item the dense molecular gas tracers coincide with the circumnuclear starburst ring, and exhibit two peaks located at or close to where the dust lanes meet this ring, and

\item the diffuse molecular gas tracers also coincide with the circumnuclear starburst ring, and $^{12}$CO~(2 -- 1) also with the innermost ends of the two dust lanes.

\end{itemize}

Using the LVG approximation, we demonstrated that:

\begin{itemize}

\item the molecular gas as observed averaged over the circumnuclear starburst ring is both relatively dense (10$^{5\:\pm\:1}$ cm$^{-3}$) and warm (well over 100~K), suggesting that massive stars have indeed formed from this gas, and

\item the diffuse molecular gas in the circumnuclear starburst ring  is unstable to gravitational collapse, providing at least one pathway for the formation of dense molecular gas.

\end{itemize}

We recreated maps of the \ion{H}{1} emission from NGC~7552 using the data taken by \citet{Dah05}, but at a higher angular resolution than those previously published.  We used this map, along with our maps in molecular lines, to derive an approximate rotation curve for NGC~7552, and computed the predicted locations of dynamical resonances under the usual assumption that the outer ends of the bar terminate at the CR.  We found that, within the statistical uncertainties, our results:

\begin{itemize}

\item are consistent with the idea that the OLR coincides closely with the outermost reaches of the twin spiral arms (which emerge from the ends of the bar) in NGC~7552, as theory predicts, and

\item show that the circumnuclear starburst ring likely lies between the iILR and oILR, as theory also predicts.

\end{itemize}

We then examined what our results imply for the formation of dense molecular gas and stars in the circumnuclear starburst ring of NGC~7552, specifically in the context of the picture proposed by \citet{Koh99} based on their observations of NGC~6951.  The results and conclusions are as follows:

\begin{itemize}

\item In both NGC~7552 and NGC~6951, emission from the dense molecular-gas tracers peaks at or close to the location where the pair of dust lanes that run along the bar meet the circumnuclear starburst ring (in the case of NGC~6951, as later improved observations by \citet{Kri07} also show).  This result suggests that the formation of dense molecular gas is promoted where the $x_1$ orbits (as closely traced by the dust lanes) and $x_2$ orbits (circumnuclear starburst ring) intersect and where collisions between gas clouds is therefore more likely, rather than further downstream along the ring as proposed by \citet{Koh99} where the velocity dispersion of $x_2$ orbits is predicted to be much smaller than that of $x_1$ orbits and hence the gas more susceptible to gravitational instabilities.  Indeed, we observe that the formation of dense molecular gas is promoted where the dust lanes meet the ring compared with regions immediately beyond along the same dust lanes.  

\item Our revised model as sketched in Figure~\ref{FIG_revised_model} also naturally explains the observed displacement between the centroids of the dense molecular gas and radio continuum knots in both NGC~7552 and NGC~6951.  This displacement ought to reflect the timescale between the formation of massive stars in the dense gas located where the dust lanes meet the circumnuclear starburst ring, and the demise of these stars as supernovae.  In the case of NGC~6951, the estimated timescale between the centroids of these knots is a few million years, at the shorter end of the 5--10~Myr timescale for the majority of massive stars formed with a Salpeter IMF to become supernovae, and in the case of NGC~7552 is an order of magnitude shorter still.  This may reflect uncertainties in the inferred angular velocities of the ring and bar, together with additional complexities mentioned below or a top-heavy IMF.

\end{itemize}

Finally, we noted that the observed properties of the circumnuclear starburst ring in NGC~7552 are more complicated than can be fully explained by the simple model proposed earlier and sketched in Figure~\ref{FIG_revised_model}.  This model can be more cleanly applied to NGC~6951, whereas the circumnuclear starburst ring of NGC~7552 exhibits not just two, but five, knots in radio continuum.  Our observations do not have sufficiently high angular resolution to spatially separate these knots in the dense (or diffuse) molecular gas tracers, so that the picture that emerges from our analyses represents a zeroth-order model for the formation of dense gas and stars in the circumnuclear starburst ring of NGC~7552.  Observations at higher angular resolution and sensitivity are necessary to tease out details not incorporated into our model, which itself is based on an important revision of the model originally proposed by \citet{Koh99} for NGC~6951.

\acknowledgements  Satoki Matsushita acknowledges grants from the National Science Council of Taiwan, NSC 100-2112-M-001-006-MY3, NSC 97-2112-M-001-021-MY3 to support Pan Hsi-An as a Masters student, and for support to conduct this work. 
The winning entry in the 2011 Australian Gemini School Astronomy Contest was submitted by Benjamin Reynolds, a student at Sutherland Shire Christian School.  We are grateful to Benjamin, Chris Onken from the Australian Gemini Office, and Dr Travis Rector from the University of Alaska, Anchorage, for making these data available to us.
We thank Dr Kotaro  Kohno for providing the line ratios of galaxies used in Figure \ref{FIG_HCN_diagram}.  PHA thanks Dr Akihiko Hirota for helping with the improvement in English.  We thank Dr Eva Schinnerer for providing image in FITS format, although not ultimately used in this manuscript.
The Submillimeter Array is a joint project between the Smithsonian Astrophysical Observatory and the Academia Sinica Institute of Astronomy and Astrophysics and is funded by the Smithsonian Institution and the Academia Sinica. The Australia Telescope is funded by the Commonwealth of Australia for operation as a National Facility managed by CSIRO.
Gemini Observatory is operated by the Association of Universities for Research in Astronomy, Inc., under a cooperative agreement with the NSF on behalf of the Gemini partnership: the National Science Foundation (United States), the Science and Technology Facilities Council (United Kingdom), the National Research Council (Canada), CONICYT (Chile), the Australian Research Council (Australia), MinistŽrio da Cincia, Tecnologia e Inova‹o (Brazil) and Ministerio de Ciencia, Tecnolog'a e Innovaci—n Productiva  (Argentina).
This research has made use of the NASA/IPAC Extragalactic Database (NED) which is operated by the Jet Propulsion Laboratory, California Institute of Technology, under contract with the National Aeronautics and Space Administration. 

{\it Facilities:} \facility{ATCA, SMA, Gemini}.

\begin{table*}
\caption{Observational parameters of radio observations.}
\begin{tabular}{l*{6}{c}r}
\hline
\hline
                                      & \ion{H}{1} &6 cm   &3 cm    &HCN(1--0)  &HCO$^{+}$(1--0)  &$^{13}$CO(2--1) &$^{12}$CO(2--1)  \\
\hline
Rest Frequency (GHz)             &  1.420        &4.786 & 8.640 & 88.631    &89.188               &  220.398         &230.538            \\
Telescope                             &  ATCA  &ATCA &ATCA   &ATCA       &ATCA                &SMA              &SMA                  \\
Projected baseline (m)            &245 -- 6000 &153 -- 6000&153 -- 6000&30 -- 200&30 -- 200&16 -- 69&16 -- 69                       \\ 
Field of view                        & 33$\arcmin$&285$\arcsec$&285$\arcsec$&36$\arcsec$&36$\arcsec$&55$\arcsec$&55$\arcsec$  \\
Velocity resolution (km s$^{-1}$)& 14             &$\dots$&$\dots$&15          &15             &18                    &       10                       \\  
Synthesized beam (arcsec$^{2}$)       &20$\times$20&2.3$\times$1.3&1.1$\times$1.1&2.6$\times$2.0 &2.6$\times$2.0&6.9$\times$2.8&7.0$\times$2.8 \\
P.A. (deg)                             &0                &0          &0       &80              &80                   &-8.9                     &-11.9            \\
Reprocessed data (yes/no)           & y              & y          & y         & n               & n                     & n                      & n                    \\
Reference                             & 1               &2           &2      & $\dots$& $\dots$& $\dots$& $\dots$                      \\                         

\hline
\label{TAB_obspara}
\end{tabular}
\footnotetext[1]{\citet{Dah05}}
\footnotetext[2]{\citet{Fo94a}}

\end{table*}

\clearpage

\clearpage

\clearpage

\clearpage


\clearpage


\clearpage

\clearpage


\clearpage

\clearpage

\clearpage

\clearpage


\clearpage
\begin{figure}
\epsscale{.70}
\plotone{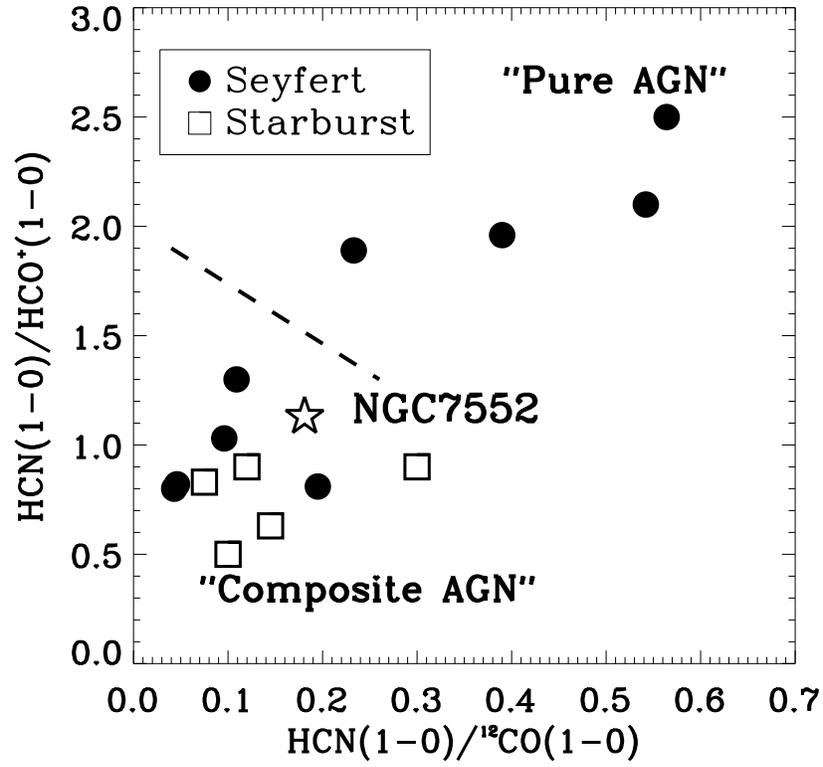}
\caption{HCN~(1 -- 0)/$^{12}$CO (1 -- 0) versus HCN~(1 -- 0)/HCO$^{+}$~(1 -- 0) integrated intensity ratios of pure Seyfert and starburst galaxies. Solid circles indicate Seyfert galaxies, open squares starburst galaxies, and an open star the position of NGC 7552 (starburst) in this plot.  Data other than for NGC 7552 are taken from \citet{Koh05}, \citet{Koh07,Koh08} and \citet{FIG_HCN_diagram}. 
\label{fig1}}
\end{figure}


\clearpage
\begin{figure}
\epsscale{.80}
\plotone{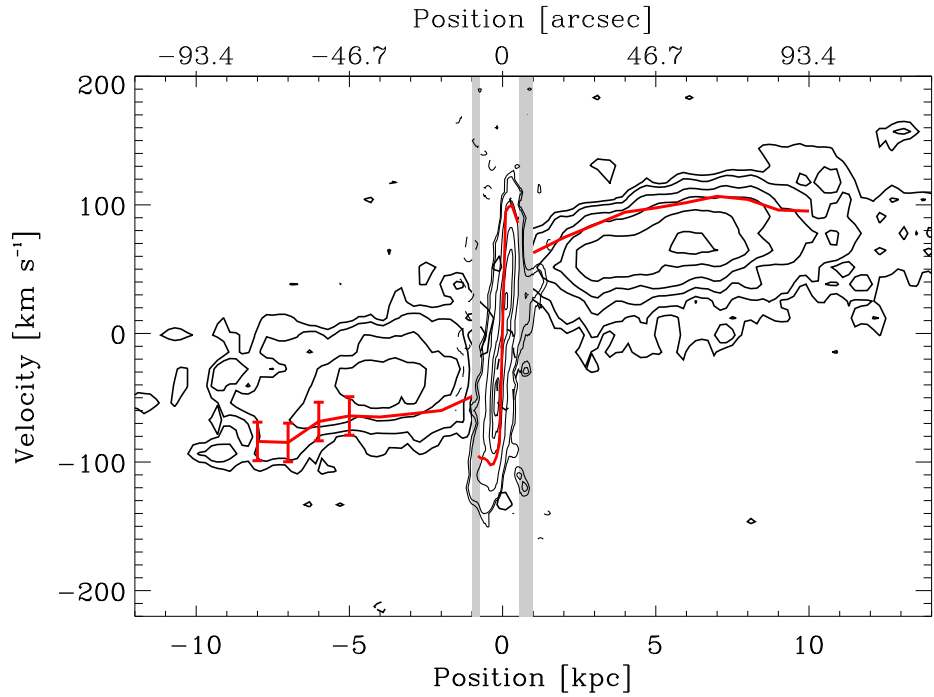}
\caption{Rotation curve, projected for an inclination of $25\degr$, shown in red curves overlaid on a P-V diagram derived from our $^{12}$CO (2 -- 1) ($-1$ to $+1$ kpc) and \ion{H}{1} maps ($\leqslant -1$ and $\geqslant 1$ kpc) at a position angle of $110\degr$ shown in contours.  Shaded regions indicate locations where the two tracers connect, and at which the rotation curve fail to track (see text). The typical projected uncertainty of the rotation curve obtained by the envelope tracing method is $\pm$15 km s$^{-1}$, and illustrated at the end of the approaching side.  Contours in $^{12}$CO (2 -- 1) are plotted in steps of -2, 2, 3, 5, 10, 20, 30, 40, and 45 $\sigma$, where 1 $\sigma$ $=$ 78 mJy beam$^{-1}$, and contours in \ion{H}{1} diagram in steps of $-2$, 2, 3, 5, 7, 10, 15, and 20$\sigma$, where 1 $\sigma$ = 1.1 mJy beam$^{-1}$.\label{FIG_PV_RC}}
\end{figure}


\clearpage
\begin{figure}
\plottwo{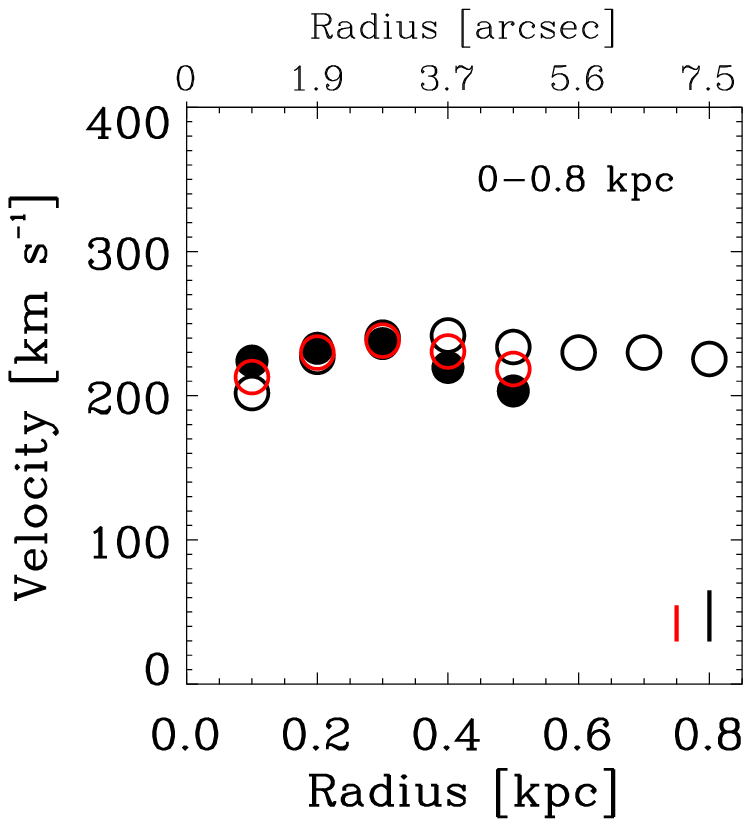}{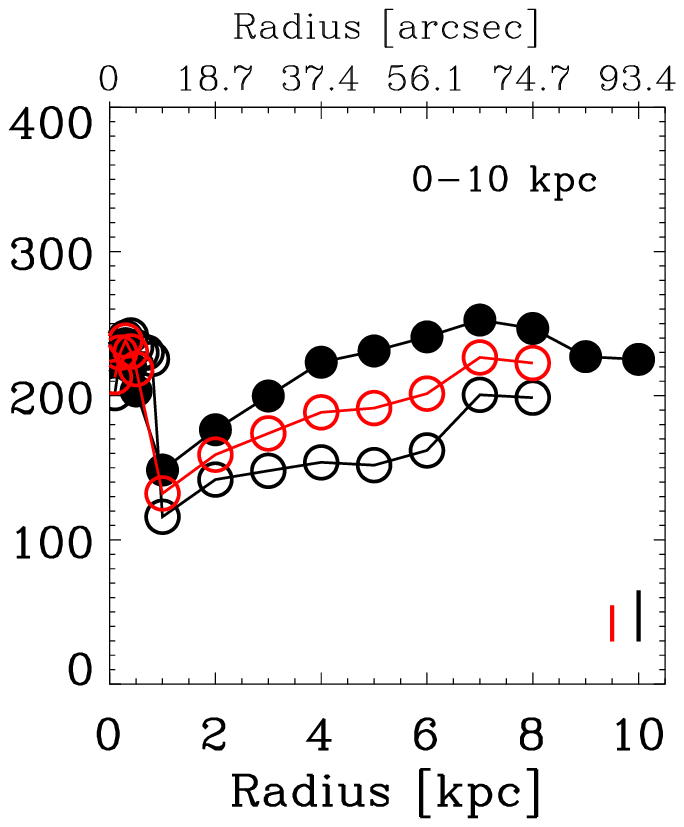}
\caption{Deprojected rotation curves derived for NGC~7552 from our $^{12}$CO~(2 -- 1) maps ($<$ 1 kpc) and \ion{H}{1} maps ($\geq$ 1 kpc).
Left panel shows the rotation curve within 1 kpc, and right panel the entire rotation curve up to 10 kpc.  Solid and open black circles denote the rotation curve of the receding and approaching sides, respectively.  We assumed a constant inclination of $25\degr$ and position angle of $110\degr$ for the kinematic major axis .  Red circles show the average rotation curve at the radii where both the receding and approaching sides can be measured. The uncertainties of the individual and average rotation curves are indicated in the lower right corner of each panel. \label{FIG_RotationCurve}}
\end{figure}


\clearpage
\begin{figure}
\epsscale{.80}
\plotone{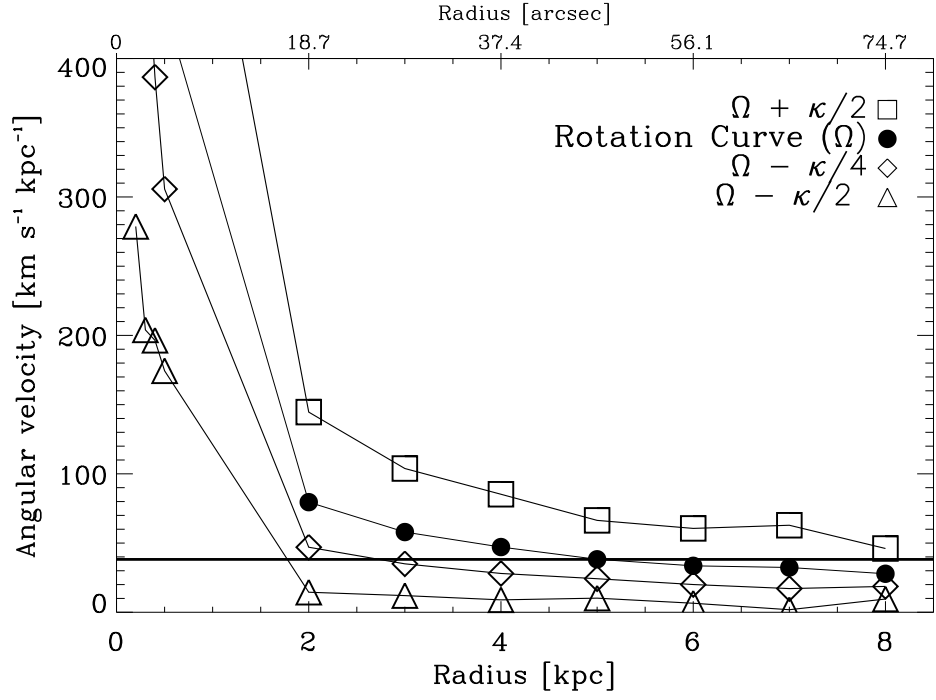}
\caption{Diagram used to infer the locations of dynamical resonances.  Black solid circles indicate the angular velocity of the galaxy, ${\Omega}$, based on an interpolation of the rotation curve shown in Figure~\ref{FIG_RotationCurve}.   Individual measurements that define the $\Omega - \kappa/2$ (ILR) curve are plotted as triangles, the $\Omega - \kappa/4$ (UHR) curve as diamonds, and the $\Omega + \kappa/2$ (OLR) curve as squares, where $\kappa$ is the epicyclic frequency of the orbit.  Resonances occur where the values of these curves are equal to the pattern speed of the bar, indicated by the thick horizontal line.
\label{FIG_Resonance}}
\end{figure}


\clearpage

\begin{figure}
\centering 
\includegraphics[scale=0.5, angle = 90]{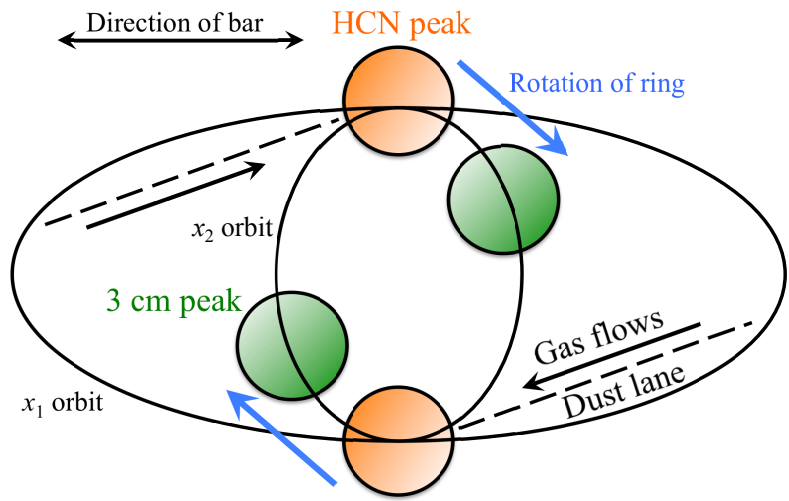}
\caption{
A sketch showing our zeroth-order picture of the formation of dense gas (HCN peaks) and synchrotron emission (3-cm peaks) resulting from supernova explosions of stars that formed from the dense gas in the circumnuclear starburst ring of NGC~7552.  This sketch is based on a revision of the picture proposed by Kohno et al. (1999) for NGC~6591.
}
\label{FIG_revised_model}
\end{figure} 

\clearpage
\begin{figure}
\epsscale{.80}
\plotone{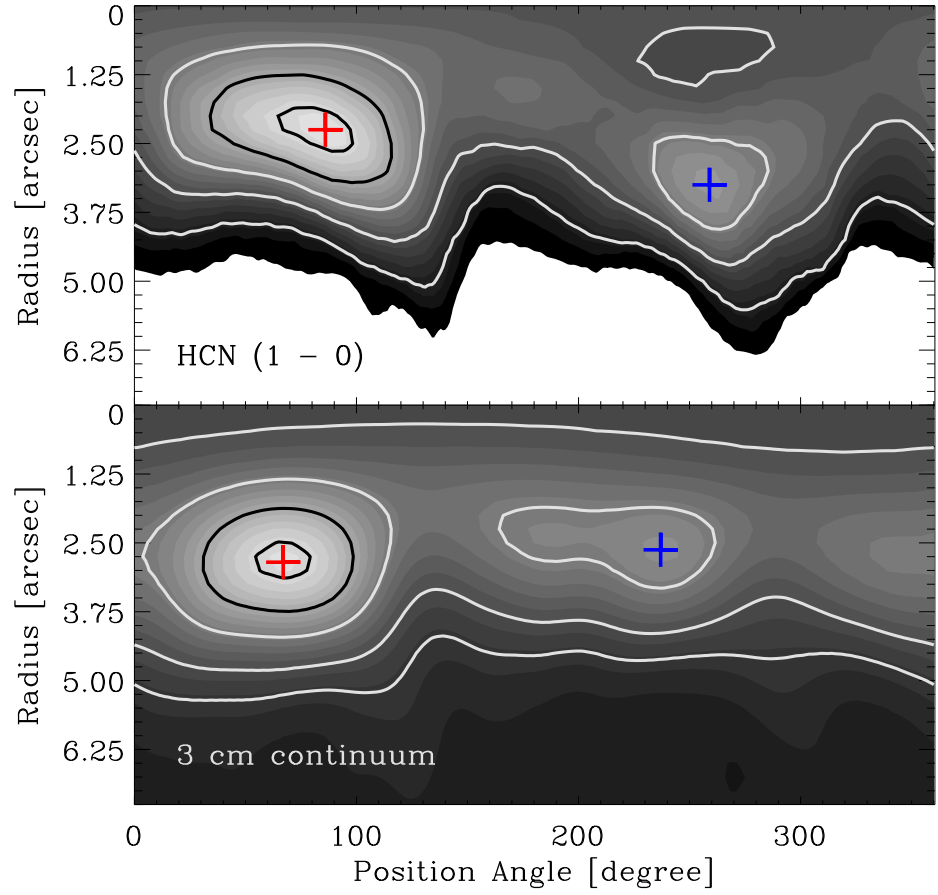}
\caption{Integrated HCN~(1 -- 0) intensity of Figure~\ref{FIG_HCN_MOM} (upper panel) and 3-cm continuum intensity of Figure~\ref{FIG_Continuum} (lower panel) plotted in polar coordinates.  The abscissa is polar angle measured from east, and the ordinate radius from center.  Crosses in both panels indicate the centroids of prominent pairs of knots in the emission.\label{FIG_Polar_Coor}}
\end{figure}


\end{document}